\documentclass[useAMS,usenatbib]{mn2e}
\usepackage{mn2e-breakabs}
\usepackage{graphicx}
\usepackage{subfigure}
\usepackage{times}
\usepackage{caption}
\usepackage{rotating}
\usepackage{rotfloat}
\usepackage{multicol}
\usepackage{array}
\usepackage{booktabs}
\usepackage{color}
\usepackage{afterpage}
\usepackage[multidot]{grffile}
\usepackage{bm}
\usepackage{float}
\usepackage{morefloats}
\usepackage{hyperref}
\usepackage{amsmath}
\usepackage{mathrsfs}
\providecommand{\adsurl}[1]{\href{#1}{ADS}}

\newcommand{\Hunit}{\,{\rm km}\,{\rm s}^{-1}\,{\rm Mpc}^{-1}}

\def\la{\mathrel{\mathpalette\fun <}}
\def\ga{\mathrel{\mathpalette\fun >}}
\def\fun#1#2{\lower3.6pt\vbox{\baselineskip0pt\lineskip.9pt
        \ialign{$\mathsurround=0pt#1\hfill##\hfil$\crcr#2\crcr\sim\crcr}}}

\newcommand{\be}{\begin{equation}}
\newcommand{\ee}{\end{equation}}
\newcommand{\ba}{\begin{eqnarray}}
\newcommand{\ea}{\end{eqnarray}}
\newcommand{\simgt}{\,\hbox{\lower0.6ex\hbox{$\sim$}\llap{\raise0.6ex\hbox{$>$}}}\,}
\newcommand{\simlt}{\,\hbox{\lower0.6ex\hbox{$\sim$}\llap{\raise0.6ex\hbox{$<$}}}\,}

\begin{document}

\title[Single-Probe Measurements from BOSS DR12]
{The Clustering of Galaxies in the Completed SDSS-III Baryon Oscillation Spectroscopic Survey:
single-probe measurements from DR12 galaxy clustering -- towards an accurate model
}

\author[Chuang et al.]{
  \parbox{\textwidth}{
 Chia-Hsun Chuang$^{1,2}$\thanks{E-mail: achuang@aip.de},
 Marcos Pellejero-Ibanez$^{3,4,1,2}$\thanks{E-mail: mpi@iac.es},
 Sergio~Rodr\'iguez-Torres$^{1,5,6}$,
 Ashley J. Ross$^{7,9}$,
 Gong-bo Zhao$^{8,9}$,
 Yuting Wang$^{8,9}$,
 Antonio J. Cuesta$^{10}$,
 J. A. Rubi\~no-Mart\'{\i}n$^{3,4}$,
Francisco Prada$^{1,6,11}$,
Shadab Alam$^{14,15}$,
Florian Beutler$^{12,9}$,
Daniel J. Eisenstein$^{13}$,
H\'ector Gil-Mar\'{i}n$^{14,15}$,
Jan Niklas Grieb$^{16,17}$,
Shirley Ho$^{18,19,12,20}$,
Francisco-Shu Kitaura$^{2,12,20}$,
Will J. Percival$^{9}$,
Graziano Rossi$^{21}$,
Salvador Salazar-Albornoz$^{16,17}$,
Lado Samushia$^{22,24,9}$,
Ariel G. S\'anchez$^{17}$,
Siddharth Satpathy$^{18,19}$,
An\v{z}e Slosar$^{24}$,
Jeremy L. Tinker$^{25}$,
Rita Tojeiro$^{26}$,
Mariana Vargas-Maga\~na$^{27}$,
Jose A Vazquez$^{24}$,
Joel R. Brownstein$^{28}$,
Robert C Nichol$^{9}$,
Matthew D Olmstead$^{29}$
}
  \vspace*{4pt} \\
$^1$ Instituto de F\'{\i}sica Te\'orica, (UAM/CSIC), Universidad Aut\'onoma de Madrid,  Cantoblanco, E-28049 Madrid, Spain \\
$^{2}$ Leibniz-Institut f\"ur Astrophysik Potsdam (AIP), An der Sternwarte 16, 14482 Potsdam, Germany\\
$^{3}$ Instituto de Astrof\'isica de Canarias (IAC), C/V\'ia L\'actea, s/n, E-38200, La Laguna, Tenerife, Spain\\
$^{4}$ Departamento Astrof\'isica, Universidad de La Laguna (ULL), E-38206 La Laguna, Tenerife, Spain\\
$^{5}$ Campus of International Excellence UAM+CSIC, Cantoblanco, E-28049 Madrid, Spain\\
$^{6}$ Departamento de F\'isica Te\'orica M8, Universidad Autonoma de Madrid (UAM), Cantoblanco, E-28049, Madrid, Spain\\
$^{7}$ Center for Cosmology and Astroparticle Physics, Department of Physics, The Ohio State University, OH 43210, USA\\ 
$^{8}$ National Astronomy Observatories, Chinese Academy of Science, Beijing, 100012, P.R.China\\
$^{9}$ Institute of Cosmology and Gravitation, University of Portsmouth, Dennis Sciama Building, Portsmouth PO1 3FX, UK\\ 
$^{10}$ Institut de Ci{\`e}ncies del Cosmos (ICCUB), Universitat de Barcelona (IEEC-UB), Mart{\'\i} i Franqu{\`e}s 1, E08028 Barcelona, Spain\\
$^{11}$ Instituto de Astrof\'{\i}sica de Andaluc\'{\i}a (CSIC), Glorieta de la Astronom\'{\i}a, E-18080 Granada, Spain \\ 
$^{12}$ Lawrence Berkeley National Lab, 1 Cyclotron Rd, Berkeley CA 94720, USA\\ 
$^{13}$ Harvard-Smithsonian Center for Astrophysics, 60 Garden St., Cambridge, MA 02138, USA\\ 
$^{14}$ Sorbonne Universités, Institut Lagrange de Paris (ILP), 98 bis Boulevard Arago, 75014 Paris, France \\
$^{15}$ Laboratoire de Physique Nucléaire et de Hautes Energies, Universit\'e Pierre et Marie Curie, 4 Place Jussieu, 75005 Paris, France \\
$^{16}$ Universit\"ats-Sternwarte M\"unchen, Scheinerstrasse 1, 81679, Munich, Germany\\ 
$^{17}$ Max-Planck-Institut f\"ur extraterrestrische Physik, Postfach 1312, Giessenbachstr., 85741 Garching, Germany\\ 
$^{18}$ McWilliams Center for Cosmology, Department of Physics, Carnegie Mellon University, 5000 Forbes Ave., Pittsburgh, PA 15213, USA\\ 
$^{19}$ The McWilliams Center for Cosmology, Carnegie Mellon University, 5000 Forbes Ave., Pittsburgh, PA 15213, USA\\ 
$^{20}$ Departments of Physics and Astronomy, University of California, Berkeley, CA 94720, USA\\ 
$^{21}$Department of Physics and Astronomy, Sejong University, Seoul, 143-747, Korea\\ 
$^{22}$ Kansas State University, Manhattan KS 66506, USA\\          
$^{23}$ National Abastumani Astrophysical Observatory, Ilia State University, 2A Kazbegi Ave., GE-1060 Tbilisi, Georgia\\  
$^{24}$ Brookhaven National Laboratory, Upton, NY 11973\\ 
$^{25}$ Center for Cosmology and Particle Physics, Department of Physics, New York University, 4 Washington Place, New York, NY 10003, USA\\ 
$^{26}$ School of Physics and Astronomy, University of St Andrews, St Andrews, KY16 9SS, UK\\ 
$^{27}$ Instituto de F\'{\i}sica, Universidad Nacional Aut\'onoma de M\'exico, Apdo. Postal 20-364, M\'exico\\ 
$^{28}$ Department of Physics and Astronomy, University of Utah, 115 S 1400 E, Salt Lake City, UT 84112, USA \\
$^{29}$ Department of Chemistry and Physics, King's College, 133 North River St, Wilkes Barre, PA 18711, USA\\ 
}

\date{\today} 

\maketitle

\begin{abstract}
We analyse the broad-range shape of the monopole
and quadrupole correlation functions of the BOSS Data Release 12 (DR12) CMASS and LOWZ galaxy sample to obtain constraints on the Hubble expansion rate $H(z)$, the angular-diameter distance $D_A(z)$, 
the normalised growth rate $f(z)\sigma_8(z)$, and the physical matter density $\Omega_mh^2$. 
We adopt wide and flat priors on all model parameters in order to ensure the results are those of a `single-probe' galaxy clustering analysis. We also marginalise over three nuisance terms that account for potential observational systematics affecting the measured monopole. The Monte Carlo Markov Chain analysis with such wide priors and additional polynomial functions is computationally expensive for advanced theoretical models.
We develop a new methodology to speed up by scanning the parameter space using a fast model first and then applying importance sampling using a slower but more accurate model.

Our measurements for DR12 galaxy sample, using the range $40h^{-1}$Mpc $<s<180h^{-1}$Mpc, are
$\{D_A(z)r_{s,fid}/r_s$, $H(z)r_s/r_{s,fid}$, $f(z)\sigma_8(z)$, $\Omega_m h^2\}$ = $\{956\pm28$ $\rm Mpc$, $75.0\pm4.0$ $\Hunit$, $0.397 \pm 0.073$, $0.143\pm0.017\}$ at $z=0.32$ and $\{1421\pm23$ $\rm Mpc$, $96.7\pm2.7$ $\Hunit$, $0.497 \pm 0.058$, $0.137\pm0.015\}$ at $z=0.59$ 
where $r_s$ is the comoving sound horizon at the drag epoch and $r_{s,fid}=147.66$ Mpc is the sound scale of the fiducial cosmology used in this 
study. 

In addition, we divide each sample (CMASS and LOWZ) into two redshift bins (four bins in total) to increase the sensitivity of redshift evolution. However, we do not find improvements in terms of constraining dark energy model parameters. Combining our measurements with Planck data, we obtain 
$\Omega_m=0.306\pm0.009$, $H_0=67.9\pm0.7\Hunit$, and $\sigma_8=0.815\pm0.009$ assuming $\Lambda$CDM; 
$\Omega_k=0.000\pm0.003$ assuming oCDM; $w=-1.01\pm0.06$ assuming $w$CDM; and $w_0=-0.95\pm0.22$ and $w_a=-0.22\pm0.63$ assuming $w_0w_a$CDM. 
Our results show no tension with the flat $\Lambda$CDM cosmological paradigm.
This paper is part of a set that analyses the final galaxy clustering dataset from BOSS.

\end{abstract}


\section{Introduction} \label{sec:intro}

The cosmic large-scale structure from galaxy redshift surveys provides a 
powerful probe of the properties of dark energy and the time dependence of any cosmological model in a manner that is highly complementary to measurements of the cosmic microwave 
background (CMB) (e.g., \citealt{Bennett:2012zja,Ade:2013sjv}), supernovae (SNe) 
\citep{Riess:1998cb,Perlmutter:1998np}, and weak lensing (see e.g. \citealt{VanWaerbeke:2003uq} for a review).

The amount of galaxy redshift 
surveys has dramatically increased in the last decades. The 2dF Galaxy Redshift Survey (2dFGRS) 
obtained 221,414 galaxy redshifts at $z<0.3$ \citep{Colless:2001gk,Colless:2003wz}, 
and the Sloan Digital Sky Survey (SDSS, \citealt{York:2000gk}) collected 
930,000 galaxy spectra in the Seventh Data Release (DR7) at $z<0.5$ \citep{Abazajian:2008wr}.
WiggleZ collected spectra of 240,000 emission-line galaxies at $0.5<z<1$ over 
1000 square degrees \citep{Drinkwater:2009sd, Parkinson:2012vd}, and the Baryon Oscillation Spectroscopic Survey (BOSS, \citealt{Dawson:2012va}) of the SDSS-III \citep{Eisenstein:2011sa} has observed 1.5 million luminous red galaxies (LRGs) at $0.1<z<0.7$ over 10,000 square degrees.
The newest BOSS data set has been made publicly available in SDSS Data Release 12 (DR12, \citealt{Alam:2015mbd}).
The planned space mission Euclid\footnote{http://sci.esa.int/euclid} will survey over 30 million emission-line galaxies at $0.7<z<2$ over 15,000 deg$^2$ 
(e.g. \citealt{Laureijs:2011gra}), and the upcoming ground-based experiment DESI\footnote{http://desi.lbl.gov/} (Dark Energy Spectroscopic Instrument) will survey 20 million galaxy redshifts up to $z=1.7$ and 
600,000 quasars ($2.2 < z < 3.5$) over 14,000 deg$^2$ \citep{Schlegel:2011zz}.
The proposed WFIRST\footnote{http://wfirst.gsfc.nasa.gov/} satellite would map 17 million galaxies in the redshift
range $1.3 < z < 2.7$ over 3400 deg$^2$, with a larger area 
possible with an extended mission \citep{Green:2012mj}.

The methodologies of the data analyses of galaxy clustering have also developed along with the growing survey volumes. The observed galaxy data have been analysed, and the cosmological results delivered, using both the power spectrum 
(see, e.g., \citealt{Tegmark:2003uf,Hutsi:2005qv,Padmanabhan:2006cia,Blake:2006kv,Percival:2007yw,Percival:2009xn,Reid:2009xm,Montesano:2011bp}), 
and the correlation function (see, e.g., 
\citealt{Eisenstein:2005su,Okumura:2007br,Cabre:2008sz,Martinez:2008iu,Sanchez:2009jq,Kazin:2009cj,Chuang:2010dv,Samushia:2011cs,Padmanabhan:2012hf,Xu:2012fw,Oka:2013cba,Hemantha:2013sea}). 
Similar analyses have been also applied to the SDSS-III BOSS \citep{Ahn:2012fh} galaxy sample 
\citep{Anderson:2012sa,Manera:2012sc,Nuza:2012mw,Reid:2012sw,Samushia:2012iq,Tojeiro:2012rp, Anderson:2013oza, Chuang:2013hya, Sanchez:2013uxa, Kazin:2013rxa,Wang:2014qoa,Anderson:2013zyy,Beutler:2013yhm,Samushia:2013yga,Chuang:2013wga,Sanchez:2013tga,Ross:2013vla,Tojeiro:2014eea,Reid:2014iaa,Alam:2015qta,Gil-Marin:2015nqa,Gil-Marin:2015sqa,Cuesta:2015mqa}.

In principle, the Hubble expansion rate $H(z)$, the angular-diameter distance $D_A(z)$, the normalized growth rate $f(z)\sigma_8(z)$, and
the physical matter density $\Omega_mh^2$ can be well constrained by analysing the galaxy clustering data alone.
\cite{Eisenstein:2005su} demonstrated the feasibility of measuring $\Omega_mh^2$ and an effective distance, $D_V(z)$, 
from the SDSS DR3 \citep{Abazajian:2004it} LRGs, where $D_V(z)$ corresponds to a combination of $H(z)$ and $D_A(z)$. 
\cite{Chuang:2011fy} measured $H(z)$ and $D_A(z)$ simultaneously using the galaxy clustering data from 
the two dimensional two-point correlation function of SDSS DR7 \citep{Abazajian:2008wr} LRGs. The methodology has been commonly known as the application of Alcock-Paczynski effect \citep{Alcock:1979mp} on large-scale structure.
The methodology has been improved and also applied to different galaxy samples, e.g., see \cite{Chuang:2012qt,Chuang:2012ad,Reid:2012sw,Blake:2012pj,Xu:2012fw}. 

Galaxy clustering allows us to differentiate between smooth dark energy and modified gravity as the cause for cosmic acceleration through the simultaneous measurements 
of the cosmic expansion history $H(z)$ and the growth rate of cosmic large scale structure, $f(z)$ \citep{Guzzo:2008ac,Wang:2007ht,Blake:2012pj}. 
However, to measure $f(z)$, one must determine the galaxy bias $b$, which requires measuring higher-order statistics of the galaxy clustering (see \citealt{Verde:2001sf}).
\cite{Song:2008qt} proposed using the normalized growth rate, $f(z)\sigma_8(z)$, which would avoid the uncertainties from the galaxy bias.
\cite{Percival:2008sh} developed a method to measure $f(z)\sigma_8(z)$ and applied it on simulations.
\cite{Wang:2012rn} estimated expected statistical constraints on
dark energy and modified gravity, including redshift-space distortions and other constraints from galaxy clustering, using a Fisher matrix formalism.
$f(z)\sigma_8(z)$ has been measured from observed data in addition to $H(z)$ and $D_A(z)$ (e.g., see \citealt{Samushia:2011cs,Blake:2012pj,Reid:2012sw,Chuang:2013hya,Wang:2014qoa,Anderson:2013zyy,Beutler:2013yhm,Chuang:2013wga,Samushia:2013yga})
 determined $f(z)\sigma_8(z)$ from the SDSS DR7 LRGs.
\cite{Blake:2012pj} measured $H(z)$, $D_A(z)$, and $f(z)\sigma_8(z)$ from the WiggleZ Dark Energy Survey galaxy sample.
Analyses have been performed to measure $H(z)$, $D_A(z)$, and $f(z)\sigma_8(z)$ from the SDSS BOSS galaxy sample \citep{Reid:2012sw,Chuang:2013hya,Wang:2014qoa,Anderson:2013zyy,Beutler:2013yhm,Chuang:2013wga,Samushia:2013yga,Alam:2015qta,Gil-Marin:2015sqa}.

In \cite{Chuang:2013wga}, we minimize the potential bias introduced by priors and observational systematics, so that one can safely combine our single-probe measurements with other data sets (i.e. CMB, SNe, etc.) to constrain the cosmological parameters of a given dark energy model. However, due to the large parameter space, the Monte Carlo Markov Chain analysis becomes expensive and makes it difficult to use more advanced/slow models. In this study, we develop a new methodology to speed up the analysis with two steps: 1) generate Marcos chains with a fast model (less accurate); 2) replace/calibrate the likelihoods with a accurate model (slower). For convenience, we use the "Gaussian streaming model" described in \cite{Reid:2011ar}, while there have been more developments, e.g. \cite{Carlson:2012bu,Wang:2013hwa,Taruya:2013my,Vlah:2013lia,White:2014gfa,Taruya:2014faa,Bianchi:2014kba,Vlah:2015sea,Okumura:2015fga}. Although the model we use might not be the most accurate model to date, it is good enough for our purpose and the scale ranges used in this study as we will demonstrate.

This paper is organized as follows. In Section \ref{sec:data}, we introduce the SDSS-III/BOSS DR12 galaxy sample and mock catalogues used 
in our study. In Section \ref{sec:method}, we describe the details of the 
methodology that constrains cosmological parameters from our galaxy clustering analysis. 
In Section \ref{sec:results}, we present our single-probe cosmological measurements. 
In Section \ref{sec:application}, given some simple dark energy models, we present the cosmological constraints from our measurements and the combination with other data sets.
We compare our results with other studies in \ref{sec:compare}.
We summarize and conclude in Section \ref{sec:conclusion}.

\section{Data sets} 
\label{sec:data}

\subsection{The CMASS and LOWZ Galaxy Catalogues}
\label{sec:galaxy}
The Sloan Digital Sky Survey (\citealt{Fukugita:1996qt,Gunn:1998vh,York:2000gk,Smee:2012wd}) mapped over one quarter 
of the sky using the dedicated 2.5 m Sloan Telescope \citep{Gunn:2006tw}.
The Baryon Oscillation Sky Survey (BOSS, \citealt{Eisenstein:2011sa, Bolton:2012hz, Dawson:2012va}) is part of the SDSS-III survey. 
It has collected the spectra and redshifts for 1.5 million galaxies, 160,000 quasars and 
100,000 ancillary targets. The Data Release 12 \citep{Alam:2015mbd} has been made publicly available\footnote{http://www.sdss3.org/}.
We use galaxies from the SDSS-III BOSS DR12 CMASS catalogue in the redshift range $0.43<z<0.75$ and LOWZ catalogue in the range $0.15<z<0.43$.
CMASS samples are selected with an approximately constant stellar mass threshold \citep{Eisenstein:2011sa}; 
LOWZ sample consists of red galaxies at $z<0.4$ from the SDSS DR8 \citep{Aihara:2011sj} image data.
We are using 800,853 CMASS galaxies and  361,775 LOWZ galaxies.
Note that the number of galaxies used in this study is slightly smaller than the one used by the \cite{Acacia} (BOSS collaboration paper for final data release) by $\sim40,000$. The difference is in the LOWZ sample used (see \citealt{Acacia} for details).
The effective redshifts of these sample are $z=0.59$ and $z=0.32$ respectively.
The details of generating these samples are described in \cite{Reid:2015gra}.
In addition, we split both CMASS and LOWZ samples into two redshift bins (four bins in total). The effective redshifts are $\{$0.24, 0.37, 0.49, 0.64$\}$; and numbers of galaxies are $\{$154367, 207408, 425612, 375241$\}$.

\subsection{Mock Catalogues}
In this study we rely on a set of 2,000 mock galaxy catalogues explicitly produced to resemble the clustering of the BOSS DR12 data. In particular we make use of the MD-Patchy BOSS DR12 mock galaxy catalogues \citep{Kitaura:2015uqa}.
These mocks are generated with the PATCHY code \citep{Kitaura:2013cwa}, which uses second-order Lagrangian perturbation theory (2LPT; e.g. see \cite{Buchert:1993ud,Bouchet:1994xp,Catelan:1994ze}) combined with a spherical collapse
model (see \citealt{Bernardeau:1993ac,Mohayaee:2005xm,Neyrinck:2012bf}) on small scales to obtain a dark matter field on a mesh (augment Lagrangian perturbation theory (ALPT); \citealt{Kitaura:2012tj}). This mesh is then populated with galaxies following a deterministic and a stochastic bias, whose parameters have been constrained to match accurately the 2- and 3-point statistics  \citep{Kitaura:2014mja}.
The calibration was performed on accurate N-body-based reference catalogues using halo abundance matching to reproduce the number density,
clustering bias, selection function, and survey geometry of the BOSS data on 10 redshift bins \citep{Rodriguez-Torres:2015vqa}.
Peculiar motions were divided into two categories in these mocks, coherent and quasi-virialised. While the first ones are obtained using ALPT, the second ones have a density dependent stochastic component which was calibrated for the 10 redshift bins.
The mock catalogues were constructed assuming $\Lambda$CDM Planck cosmology with \{$\Omega_{\rm M}=0.307115, \Omega_{\rm b}=0.048206,\sigma_8=0.8288,n_s=0.96$\}, and a Hubble constant ($H_0=100\,h\Hunit$) given by  $h=0.6777$. 
As shown in a mock catalogue comparison study (\citealt{Chuang:2014toa}), PATCHY mocks are accurate within 5\% on scales larger than 5 Mpc/h (or $k$ smaller than 0.5 h/Mpc in Fourier space) for monopole and within 10-15\% for quadrupole. \cite{Kitaura:2015uqa} had also demonstrated the accuracy of BOSS PATCHY mock catalogues which are in very good agreement with the observed data in terms of 2- and 3-point statistics. 
These mocks have been used in recent galaxy clustering studies \citep{Cuesta:2015mqa,Gil-Marin:2015nqa,Gil-Marin:2015sqa,Rodriguez-Torres:2015vqa,Slepian:2015hca} and void clustering studies \citep{Kitaura:2015ubm,Liang:2015oqc}.  
They are also used in 
\cite{Acacia} (BOSS collaboration paper for final data release) and its companion papers including this paper and 
\cite{Ross16, 
Vargas-Magana16, 
Beutler16b, 
Satpathy16, 
Beutler16c, 
Sanchez16a, 
Grieb16, 
Sanchez16b, 
Pellejero-Ibanez16, 
Slepian16a, 
Slepian16b, 
Salazar-Albornoz16, 
Zhao16, 
Wang16}.

\section{Methodology} 
\label{sec:method}

In this section, we describe the measurement of the multipoles of the correlation function
from the observational data, construction of the theoretical prediction, 
and the likelihood analysis that leads to constraining 
cosmological parameters and dark energy. 

\subsection{Two-Dimensional Two-Point Correlation Function}
We convert the measured redshifts of the BOSS CMASS and LOWZ galaxies to comoving distances 
by assuming a fiducial model, i.e., flat $\Lambda$CDM with $\Omega_m=0.307115$ and $h=0.6777$ 
which is the same model adopted for constructing the mock catalogues (see \citealt{Kitaura:2015uqa}). 
We use the two-point correlation function estimator given by 
\cite{Landy:1993yu}:
\begin{equation}
\label{eq:xi_Landy}
\xi(s,\mu) = \frac{DD(s,\mu)-2DR(s,\mu)+RR(s,\mu)}{RR(s,\mu)},
\end{equation}
where $s$ is the separation of a pair of objects and $\mu$ is the cosine of the angle between the directions between the line of sight (LOS) and the line connecting the pair the objects. DD, DR, and RR represent the normalized data-data,
data-random, and random-random pair counts, respectively, for a given
distance range. The LOS is defined as the direction from the observer to the 
centre of a galaxy pair. Our bin size is
$\Delta s=1 \, h^{-1}$Mpc and $\Delta \mu=0.01$. 
The Landy and Szalay estimator has minimal variance for a Poisson
process. The random catalogue is generated with the radial and angular selection function of the observed galaxies. One can reduce the shot noise due
to random data by increasing the amount of random points. The number
of random data we use is about 50 times that of the observed galaxies. While
calculating the pair counts, we assign to each data point a radial
weight of $1/[1+n(z)\cdot P_w]$, where $n(z)$ is the radial
number density and $P_w = 1\cdot 10^4$ $h^{-3}$Mpc$^3$ (see  
\citealt{Feldman:1993ky}).
We include the combination of the observational weights assigned for each galaxy by
\begin{equation}\label{eq:weight}
w_{tot,i} = w_{sys,i}*(w_{rf,i}+w_{fc,i}-1),
\end{equation}
where $w_{tot,i}$ is the final weight to assign on a galaxy $i$; $w_{sys,i}$ is for removing the correlation between CMASS galaxies and both stellar density and seeing; $w_{rf,i}$ and $w_{fc,i}$ correct for missing objects due to the redshift failure and fiber collision. The details are described in \cite{Reid:2015gra} (see also \citealt{Ross:2012qm}). Later, we will also test the impact of systematics by removing $w_{sys,i}$ from the analysis.

\subsection{Multipoles of the Two-Point Correlation Function}  \label{sec:multipoles}

The traditional multipoles of the two-point correlation function, in redshift space, are defined by
\ba
\label{eq:multipole_1}
\xi_l(s) &\equiv & \frac{2l+1}{2}\int_{-1}^{1}{\rm d}\mu\, \xi(s,\mu) P_l(\mu),
\ea
where $P_l(\mu)$ is the Legendre Polynomial ($l=$0 and 2 here). 
We integrate over a spherical shell with radius $s$,
while actual measurements of $\xi(s,\mu)$ are done in discrete bins.
To compare the measured $\xi(s,\mu)$ and our theoretical model, the last integral in Eq.(\ref{eq:multipole_1}) should be converted into a sum,
\begin{equation}\label{eq:multipole}
 \hat{\xi}_l(s) \equiv \frac{\displaystyle\sum_{s-\frac{\Delta s}{2} < s' < s+\frac{\Delta s}{2}}\displaystyle\sum_{0\leq\mu\leq1}(2l+1)\xi(s',\mu)P_l(\mu)}{\mbox{Number of bins used in the numerator}},
\end{equation}
where $\Delta s=5$ $h^{-1}$Mpc in this work.

\begin{figure*}
\centering
\subfigure{\includegraphics[width=1 \columnwidth,clip,angle=-0]{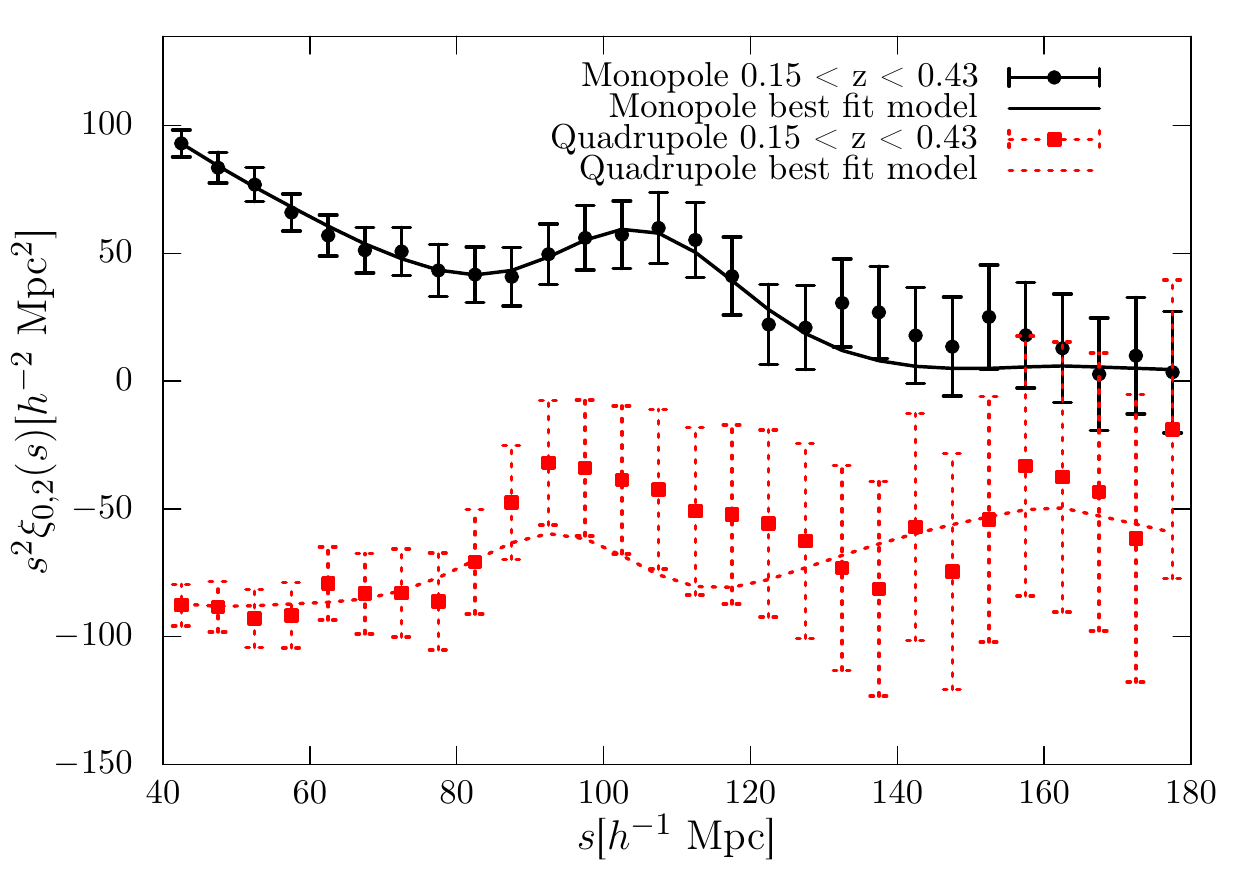}}
\subfigure{\includegraphics[width=1 \columnwidth,clip,angle=-0]{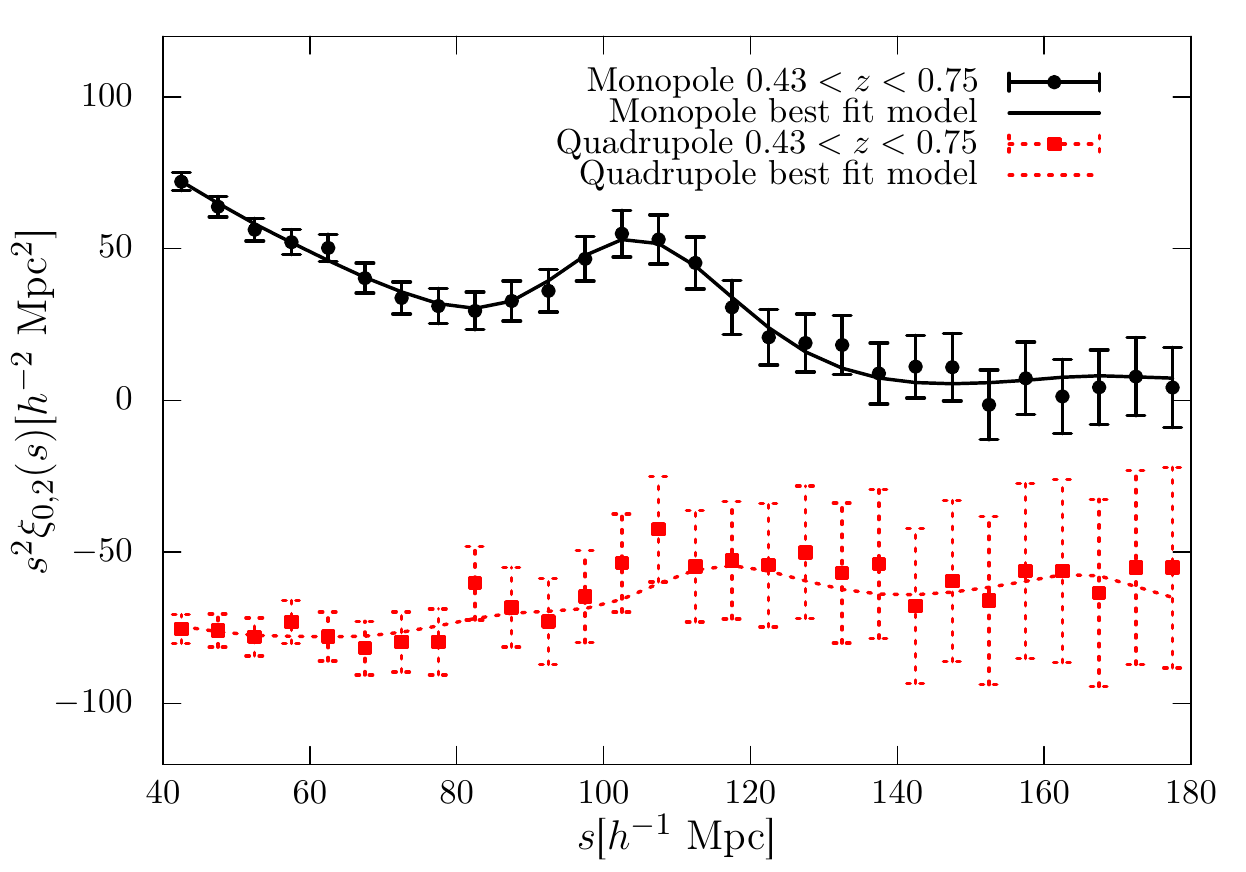}}
\caption{Measurement of monopole and quadrupole of the correlation function from two redshift bins.
Left panel: measurements from the BOSS DR12 LOWZ galaxy sample within $0.15<z<0.43$ compared to the best-fitting theoretical models  (solid lines). 
The $\chi^2$ per degree of freedom (d.o.f.) is 0.91.
Right panel: measurements from the BOSS DR12 CMASS galaxy sample within $0.43<z<0.75$ compared to the best-fitting theoretical models  (solid lines).
The $\chi^2$/d.o.f. is 1.07.
The error bars are the square root of the diagonal elements of the covariance matrix.
In this study, our fitting scale ranges are $40h^{-1}$Mpc $<s<180h^{-1}$Mpc; the bin size is $5h^{-1}$Mpc.
}
\label{fig:mp_cmass_lowz}
\end{figure*}

\begin{figure*}
\centering
\subfigure{\includegraphics[width=1 \columnwidth,clip,angle=-0]{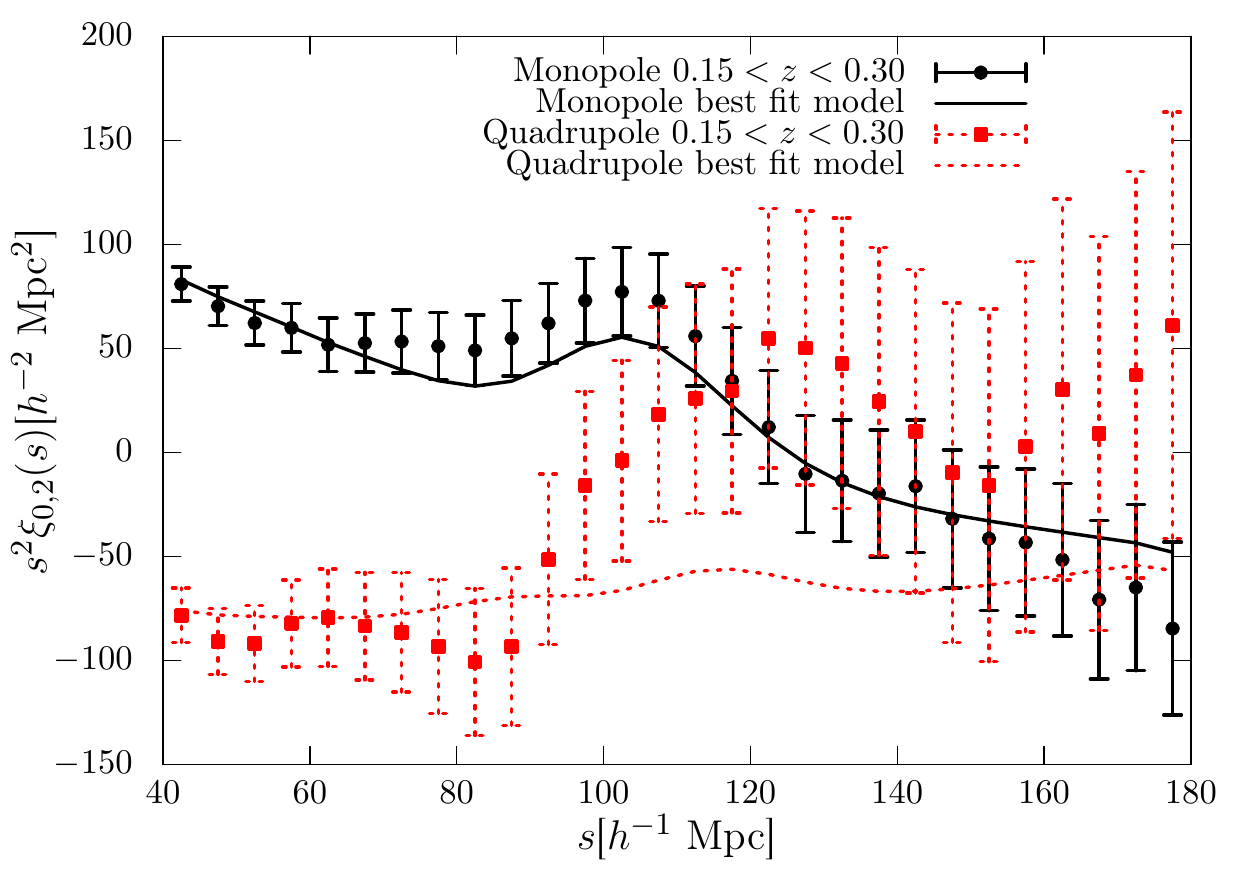}}
\subfigure{\includegraphics[width=1 \columnwidth,clip,angle=-0]{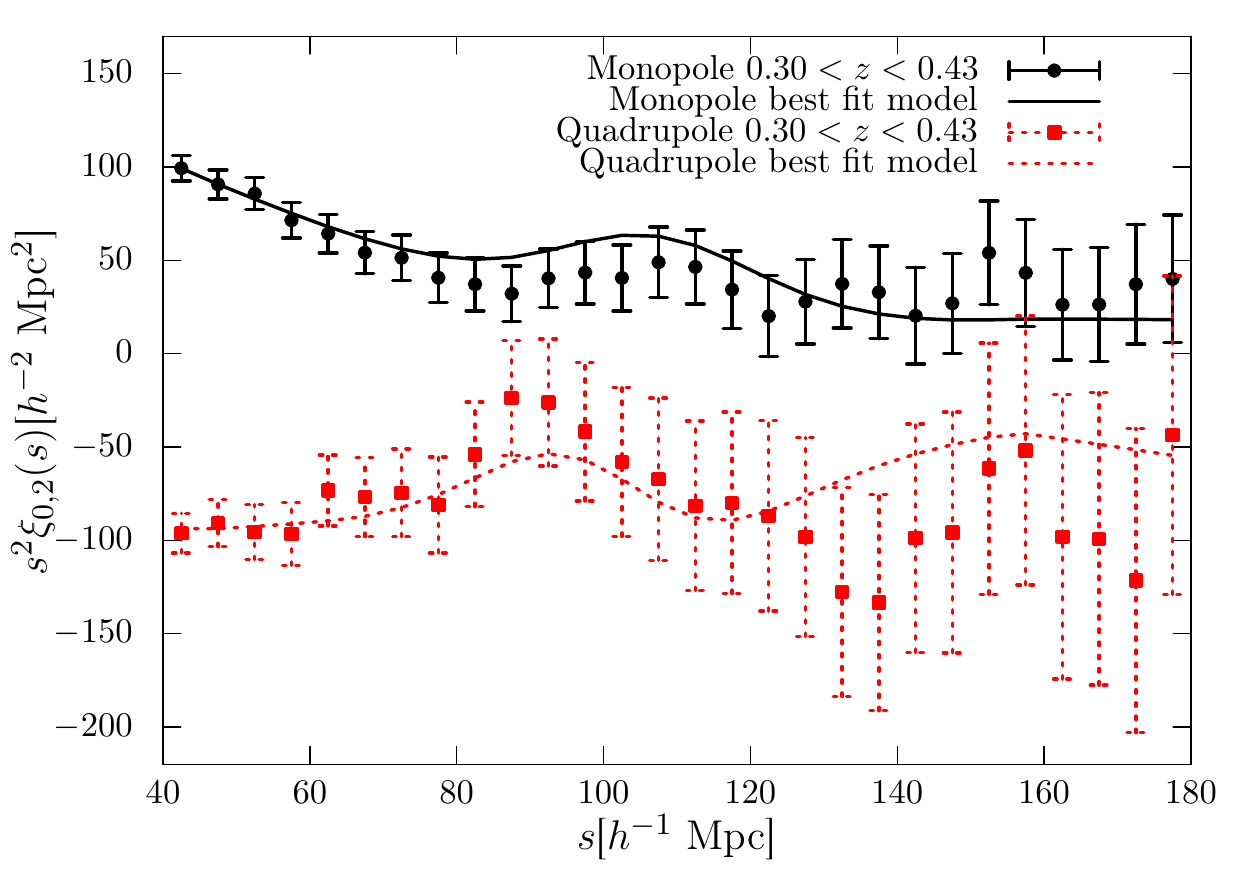}}
\subfigure{\includegraphics[width=1 \columnwidth,clip,angle=-0]{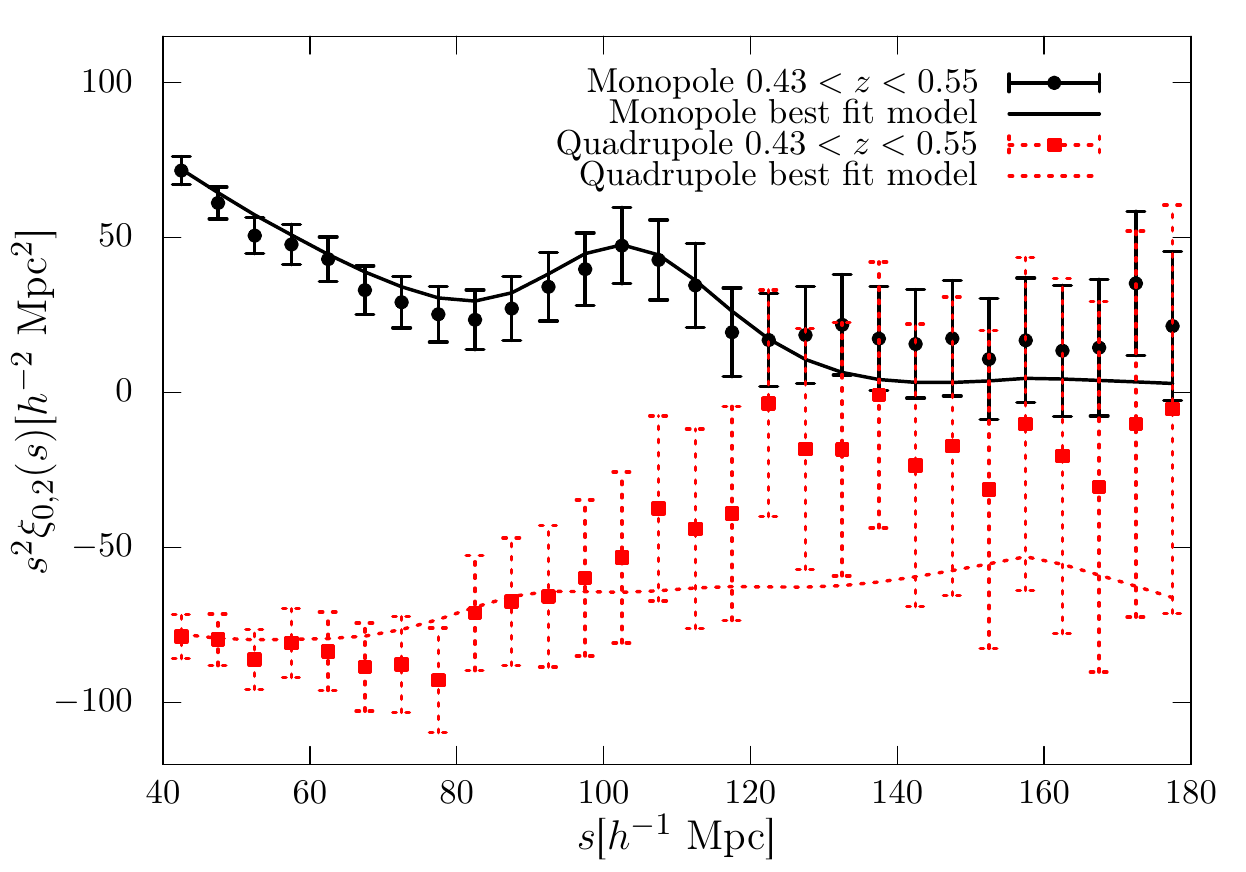}}
\subfigure{\includegraphics[width=1 \columnwidth,clip,angle=-0]{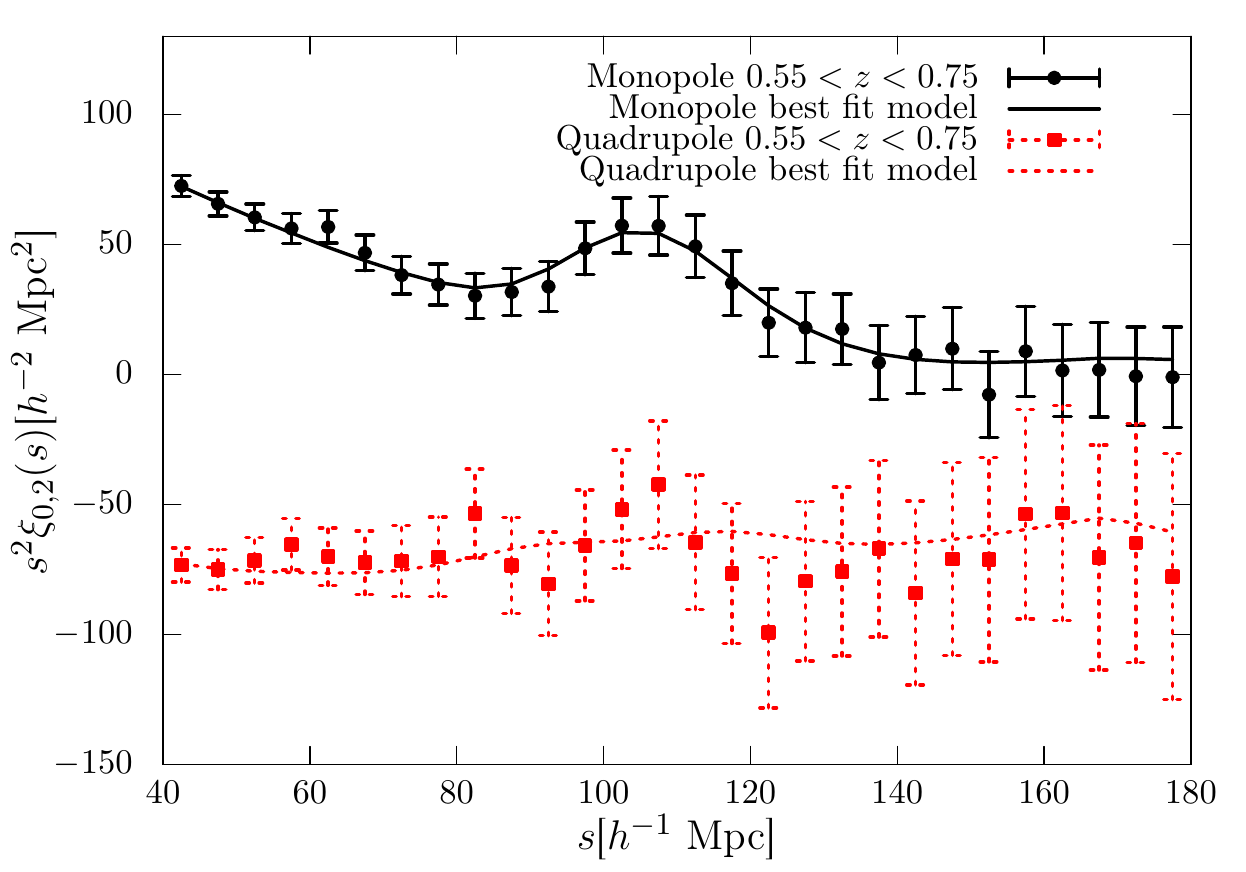}}
\caption{Measurement of monopole and quadrupole of the correlation function from four redshift bins.
Top left panel: measurements from the BOSS DR12 LOWZ galaxy sample within $0.15<z<0.30$ compared to the best-fitting theoretical models  (solid lines).
The $\chi^2$/d.o.f. is 0.71.
Top right panel: measurements from the BOSS DR12 LOWZ galaxy sample within $0.30<z<0.43$ compared to the best-fitting theoretical models  (solid lines).
The $\chi^2$/d.o.f. is 1.20.
Bottom left panel: measurements from the BOSS DR12 CMASS galaxy sample within $0.43<z<0.55$ compared to the best-fitting theoretical models  (solid lines).
The $\chi^2$/d.o.f. is 1.15.
Bottom right panel: measurements from the BOSS DR12 CMASS galaxy sample within $0.55<z<0.75$ compared to the best-fitting theoretical models  (solid lines).
The $\chi^2$/d.o.f. is 0.85.
The error bars are the square root of the diagonal elements of the covariance matrix.
In this study, our fitting scale ranges are $40h^{-1}$Mpc $<s<180h^{-1}$Mpc; the bin size is $5h^{-1}$Mpc.
}
\label{fig:mp_4bins}
\end{figure*}

Fig.\ref{fig:mp_cmass_lowz} shows the monopole ($\hat{\xi}_0$) and quadrupole ($\hat{\xi}_2$) measured from the BOSS CMASS and LOWZ galaxy sample compared with the best fit theoretical models. We split both CMASS and LOWZ sample into two redshift bins and show the multipoles from these four bins in Fig. \ref{fig:mp_4bins}.

We are using the scale range $s=40-180\,h^{-1}$Mpc and the bin size is 5 $h^{-1}$Mpc. 
Fig.~\ref{fig:mp_cmass_lowz} and\ref{fig:mp_4bins} show the measured multipoles from various redshift ranges and their best fits. The theoretical models will be described in the next section.

\subsection{Theoretical Two-Point Correlation Function}
\label{sec:model_2PCF}
We use two theoretical models for this study. One is the two-dimensional dewiggle model \citep{Eisenstein:2006nj} and the other is the Gaussian streaming model \citep{Reid:2011ar}. The former model is very fast but less accurate for high bias tracers; the latter is more accurate but much slower in terms of computation. We develop a new methodology to take the advantages from both of them.

\subsubsection{Fast model -- two-dimensional dewiggle model}
We use the fast model (two-dimensional dewiggle model) which includes the linear bias, nonlinear evolution at BAO scales, linear redshift space distortion, and nonlinear redshift space distortion at BAO scales on top of the linear theoretical model.
The theoretical model can be constructed by first and higher order perturbation theory. 
The procedure of constructing our fast model in redshift space is the following:
First, we adopt the cold dark matter model and the simplest inflation model (adiabatic initial condition).
Thus, we can compute the linear matter power spectra, $P_{lin}(k)$, by using CAMB (Code for Anisotropies in the Microwave Background, \citealt{Lewis:1999bs}). The linear power spectrum can be decomposed into two parts:
\begin{equation} \label{eq:pk_lin}
P_{lin}(k)=P_{nw}(k)+P_{BAO}^{lin}(k),
\end{equation}
where $P_{nw}(k)$ is the ``no-wiggle'' or pure CDM power spectrum calculated using Eq.(29) from \cite{Eisenstein:1997ik}. $P_{BAO}^{lin}(k)$ is the ``wiggled'' part defined by Eq. (\ref{eq:pk_lin}).
The nonlinear damping effect of the ``wiggled'' part, in redshift space, can be well approximated following \cite{Eisenstein:2006nj} by
\begin{equation} \label{eq:bao}
P_{BAO}^{nl}(k,\mu_k)=P_{BAO}^{lin}(k)\cdot \exp\left(-\frac{k^2}{2k_\star^2}[1+\mu_k^2(2f+f^2)]\right),
\end{equation}
where $\mu_k$ is the cosine of the angle between ${\bf k}$ and the LOS, $f$ is the growth rate, and
$k_\star$ is computed following \cite{Crocce:2005xz} and \cite{Matsubara:2007wj} by
\begin{equation} \label{eq:kstar}
k_\star=\left[\frac{1}{3\pi^2}\int P_{lin}(k)dk\right]^{-1/2}.
\end{equation}
The dewiggled power spectrum is
\begin{equation} \label{eq:pk_dw}
P_{dw}(k,\mu_k)=P_{nw}(k)+P_{BAO}^{nl}(k,\mu_k).
\end{equation}

Besides the nonlinear redshift distortion introduced above, we include the linear redshift distortion as follows in order to
obtain the galaxy power spectrum in redshift space at large scales \citep{Kaiser:1987qv},
\begin{eqnarray} \label{eq:pk_2d}
P_g^s(k,\mu_k)&=&b^2(1+\beta\mu_k^2)^2P_{dw}(k,\mu_k),
\end{eqnarray}
where $b$ is the linear galaxy bias and $\beta=f/b$ is the linear redshift distortion parameter.

We compute the theoretical two-point correlation 
function, $\xi(s,\mu)$, by Fourier transforming the non-linear power spectrum
$P_g^s(k,\mu_k)$. This task is efficiently performed by using Legendre polynomial expansions and one-dimensional integral convolutions as introduced in \cite{Chuang:2012qt}.

The purpose of using fast model is to mimic the slow model in a very efficient way. We thus define the following calibration functions to the fast model:
\begin{eqnarray}
\xi_0^{cal}(s)=(1-e^{-\frac{s}{s_{1}}}+e^{-\left(\frac{s}{s_{2}}\right)^2})\xi_0(s),\\
\xi_2^{cal}(s)=(1-e^{-\frac{s}{s_{3}}}+e^{-\left(\frac{s}{s_{4}}\right)^2})\xi_2(s),
\end{eqnarray}
where we find the calibration parameters, $s_1=12$, $s_2=14$, $s_3=20$, and $s_4=27h^{-1}$Mpc, by comparing the fast and slow models from a visual inspection.
Later, we will explain that the calibration parameters will speed up the convergence but will not bias the results when doing a MCMC analysis. Therefore, it is not critical to find the optimal form of calibration function and its parameters.

\subsubsection{Slow model -- Gaussian streaming model}
We use an advanced model called "Gaussian streaming model" described in
\cite{Reid:2011ar}. 
The model assumes the pairwise velocity probability distribution function is
 Gaussian and can be used to relate real space clustering and pairwise
velocity statistics of halos to their clustering in redshift space by  
\begin{eqnarray}
  1+\xi^{s}_{\rm g}(r_{\sigma},r_{\pi}) = \nonumber
\end{eqnarray}
\begin{equation}
\int \left[1+\xi^{r}_{\rm g}(r)\right]
  e^{-[r_\pi - y - \mu v_{12}(r)]^2/2\sigma_{12}^2(r,\mu)} \frac{dy}{\sqrt{2\pi\sigma^2_{12}(r,\mu)}} \label{streaming},
\end{equation}
\noindent
where $r_\sigma$ and $r_\pi$ are the redshift space transverse and LOS distances
between two objects with respect to the observer, $y$ is the {\em real} space
LOS pair separation, $\mu = y/r$, $\xi_{\rm g}^{\rm r}$ and $\xi_{\rm g}^{\rm s}$ are the real and redshift space galaxy correlation functions respectively, $v_{12}(r)$ is the average infall velocity of
galaxies separated by real-space distance $r$, and $\sigma_{12}^2(r,\mu)$ is the
rms dispersion of the pairwise velocity between two galaxies separated with
transverse (LOS) real space separation $r_{\sigma}$ ($y$).  
$\xi_{\rm g}^{\rm r}(r)$, $v_{12}(r)$ and $\sigma_{12}^2(r,\mu)$ are computed
in the framework of Lagrangian ($\xi^{\rm r}$) and standard perturbation
theories ($v_{12}$, $\sigma_{12}^2$).  

For large scales, only one nuisance parameter is necessary to describe the clustering of a
sample of halos or galaxies in this model: $b_{1L} = b-1$, the first-order
Lagrangian host halo bias in {\em real} space.
One would need another parameter, $\sigma^2_{\rm FoG}$, to model an
additive, isotropic velocity dispersion accounting for small-scale motions
of halos and galaxies (one halo term).
However, in this study, we consider relative large scales (i.e. $40 < s < 180 h^{-1}$Mpc), so that we do not include this parameter.
Further details of the model, its numerical implementation, and its accuracy
can be found in \cite{Reid:2011ar}. 

\subsubsection{Model for observational systematic errors}
It is well known that the observations could be contaminated by systematic effects, e.g. see \cite{Ross:2012qm} and Ross et al. (2016; companion paper).
To obtain robust and conservative measurements, we include a model for systematics.
The model is a simple polynomial given by
\begin{equation} \label{eq:systematic}
 A(s)=a_0+\frac{a_1}{s}+\frac{a_2}{s^2},
\end{equation}
where $a_0$, $a_1$, and $a_2$ are nuisance parameters. 
Following \cite{Chuang:2013wga}, we only include the systematics model for the monopole of the correlation function since the quadrupole is insensitive to the systematics effects of which we are aware. On the other hand, if we add another polynomial to the quadrupole as it is usually done in papers of measuring BAO only (e.g. \citealt{Anderson:2013zyy,Cuesta:2015mqa}), we would not be able to extract any information from redshift space distortions.  

\subsection{Covariance Matrix} \label{sec:covar}

We use the 2000 mock catalogues created by \citealt{Kitaura:2015uqa}
for the BOSS DR12 CMASS and LOWZ galaxy sample
to estimate the covariance matrix of the observed correlation function. 
We calculate the multipoles of the correlation functions 
of the mock catalogues and construct the covariance matrix as
\begin{equation}
 C_{ij}=\frac{1}{(N-1)(1-D)}\sum^N_{k=1}(\bar{X}_i-X_i^k)(\bar{X}_j-X_j^k),
\label{eq:covmat}
\end{equation}
where
\begin{equation}
 D=\frac{N_b +1}{N-1},
\label{eq:D}
\end{equation}
$N$ is the number of the mock catalogues, $N_b$ is the number of data bins, $\bar{X}_m$ is the
mean of the $m^{th}$ element of the vector from the mock catalogue multipoles, and
$X_m^k$ is the value in the $m^{th}$ elements of the vector from the $k^{th}$ mock
catalogue multipoles. 
We are using the scale range $s=40-180\,h^{-1}$Mpc and the bin size is 5 $h^{-1}$Mpc. 
The data points from the multipoles in the scale range considered are combined to form a 
vector, $X$, i.e.,
\be\label{eq:X}
{\bf X}=\{\hat{\xi}_0^{(1)}, \hat{\xi}_0^{(2)}, ..., \hat{\xi}_0^{(N)}; 
\hat{\xi}_2^{(1)}, \hat{\xi}_2^{(2)}, ..., \hat{\xi}_2^{(N)};...\},
\ee
where $N$ is the number of data points in each measured multipole; here $N=28$ is the same for all the redshift bins.
The length of the data vector ${\bf X}$ depends on the number of multipoles used. 
We also include the correction, $D$, introduced by \cite{Hartlap:2006kj}. 

\subsection{Likelihood}
The likelihood is taken to be proportional to $\exp(-\chi^2/2)$ \citep{press92}, 
with $\chi^2$ given by
\begin{equation} \label{eq:chi2}
 \chi^2\equiv\sum_{i,j=1}^{N_{X}}\left[X_{th,i}-X_{obs,i}\right]
 C_{ij}^{-1}
 \left[X_{th,j}-X_{obs,j}\right]
\end{equation}
where $N_{X}$ is the length of the vector used, 
$X_{th}$ is the vector from the theoretical model, and $X_{obs}$ 
is the vector from the observed data.

As explained in \cite{Chuang:2011fy}, instead of recalculating the observed correlation function while 
computing for different models, we rescale the theoretical correlation function to avoid rendering the $\chi^2$ values arbitrary (the amount of information from data sample used needs to be fixed when computing $\chi^2$).
This approach can be considered as an application of Alcock-Paczynski effect \citep{Alcock:1979mp}.
The rescaled theoretical correlation function is computed from
\begin{equation} \label{eq:inverse_theory_2d}
 T^{-1}(\xi_{th}(\sigma,\pi))=\xi_{th}
 \left(\frac{D_A(z)}{D_A^{fid}(z)}\sigma,
 \frac{H^{fid}(z)}{H(z)}\pi\right),
\end{equation}
where $\xi_{th}$ is the theoretical model described in Sec. \ref{sec:model_2PCF}, and $\chi^2$ can be rewritten as
\ba 
\label{eq:chi2_2}
\chi^2 &\equiv&\sum_{i,j=1}^{N_{X}}
 \left\{T^{-1}X_{th,i}-X^{fid}_{obs,i}\right\}
 C_{ij}^{-1} \cdot \nonumber\\
 & & \cdot \left\{T^{-1}X_{th,j}-X_{obs,j}^{fid}\right\};
\ea
where $T^{-1}X_{th}$ is the vector computed from eq.\ (\ref{eq:multipole}) from the rescaled theoretical correlation function, eq. (\ref{eq:inverse_theory_2d}).
$X^{fid}_{obs}$ is the vector from observed data measured with the fiducial model (see \citealt{Chuang:2011fy} for more details regarding the rescaling method).

\subsection{Markov Chain Monte-Carlo Likelihood Analysis} 
\label{sec:mcmc}

\subsubsection{Basic procedure}
We perform Markov Chain Monte-Carlo
likelihood analyses using CosmoMC \citep{Lewis:2002ah}. 
The parameter space that we explore spans the parameter set of
$\{H(z)$, $D_A(z)$, $\Omega_mh^2$, $\beta(z)$, $b\sigma_8(z)$, $\Omega_bh^2$, $n_s$, $b(z)$, $a_0$, $a_1$, $a_2\}$. 
The quantities $\Omega_m$ and $\Omega_b$ are the matter and
baryon density fractions, $n_s$ is the power-law index of the primordial matter power spectrum, 
$h$ is the dimensionless Hubble
constant ($H_0=100h$ km s$^{-1}$Mpc$^{-1}$), and $\sigma_8(z)$ is the normalization of the power spectrum.
The linear redshift distortion parameter can be expressed as $\beta(z)=f(z)/b$.
Thus, one can derive $f(z)\sigma_8(z)$ from the measured $\beta(z)$ and $b\sigma_8(z)$.
Among these parameters, only $\{H(z)$, $D_A(z)$, $\Omega_mh^2$, $\beta(z)$, $b\sigma_8(z)\}$ are well constrained using
the BOSS galaxy sample alone in the scale range of interest. We marginalize over the other six parameters, 
$\{\Omega_bh^2$, $n_s$, $b$, $a_0$, $a_1$, $a_2\}$, assuming a flat prior over the range 
$\{(0.01877, 0.02537)$, $(0.8676, 1.0556)$, $(1.5, 2.5)$, $(-0.003, 0.003)$, $(-3, 3)$, $(-20, 20)\}$ respectively, 
where the flat priors on $\Omega_b h^2$ and $n_s$ are centred on 
the Planck measurements with a width of 
$\pm10\sigma_{Planck}$ ($\sigma_{Planck}$ is taken from \citealt{Ade:2013zuv}). 
These priors
are sufficiently wide to ensure that CMB constraints are not double counted 
when our results are combined with CMB data \citep{Chuang:2010dv}.

\subsubsection{Generate/calibrate Markov chains with fast/slow model}
We first use the fast model (2D dewiggle model) to compute the likelihood, $\mathcal{L}_{fast}$ and generate the Markov chains. 
This step will make many trials (keep or throw away based on the MCMC algorithm) and eventually provides the chains of parameter points describing the parameter constraints and exclude the low likelihood regions of the parameter space.

Once we have the chains generated using the fast model, we modify the weight of each point in the chains by
\begin{equation}
\mathcal{W}_{new}=\mathcal{W}_{old}\frac{\mathcal{L}_{slow}}{\mathcal{L}_{fast}},
\end{equation}
where $\mathcal{L}_{slow}$ and $\mathcal{L}_{fast}$ are the likelihoods for a given point of input parameters in the chains and $\mathcal{W}_{old}$ is the original weight of the given point. We save time by computing only the "important" points without computing the likelihood of a point which we will not include eventually. The methodology is known as "importance sampling". However, the typical application of the importance sampling method is to add a likelihood from some additional data set to a given set of chains, but in this study, we will use it to replace the likelihood of a data set with a more accurate version.

It takes about 9 hours to find the best fit value using CosmoMC (i.e. action=2) with the slow model and 30 minutes (20 times faster) with the fast model.

On the scales we use for comparison with the BOSS galaxy data, the theoretical correlation 
function only depends on cosmic curvature and dark energy through the
parameters $H(z)$, $D_A(z)$, $\beta(z)$, and $b\sigma_8(z)$        
assuming that dark energy perturbations are unimportant (valid in the simplest dark energy models).
Thus we are able to extract constraints from clustering data that are independent of dark energy. 


\section{Results} \label{sec:results}

\subsection{Validate the Methodology using Mock Catalogues}
\begin{figure*}
\centering
\subfigure{\includegraphics[width=1 \columnwidth,clip,angle=-0]{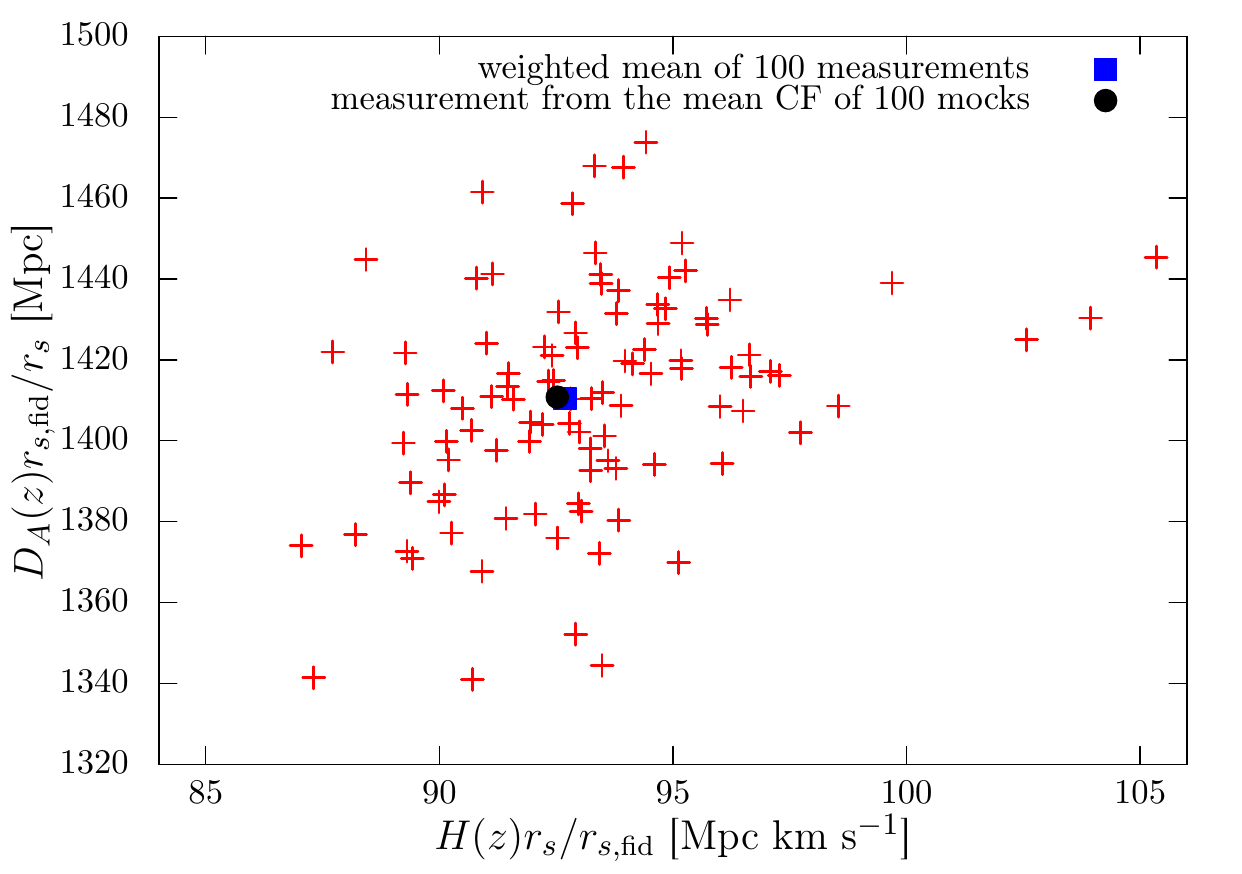}}
\subfigure{\includegraphics[width=1 \columnwidth,clip,angle=-0]{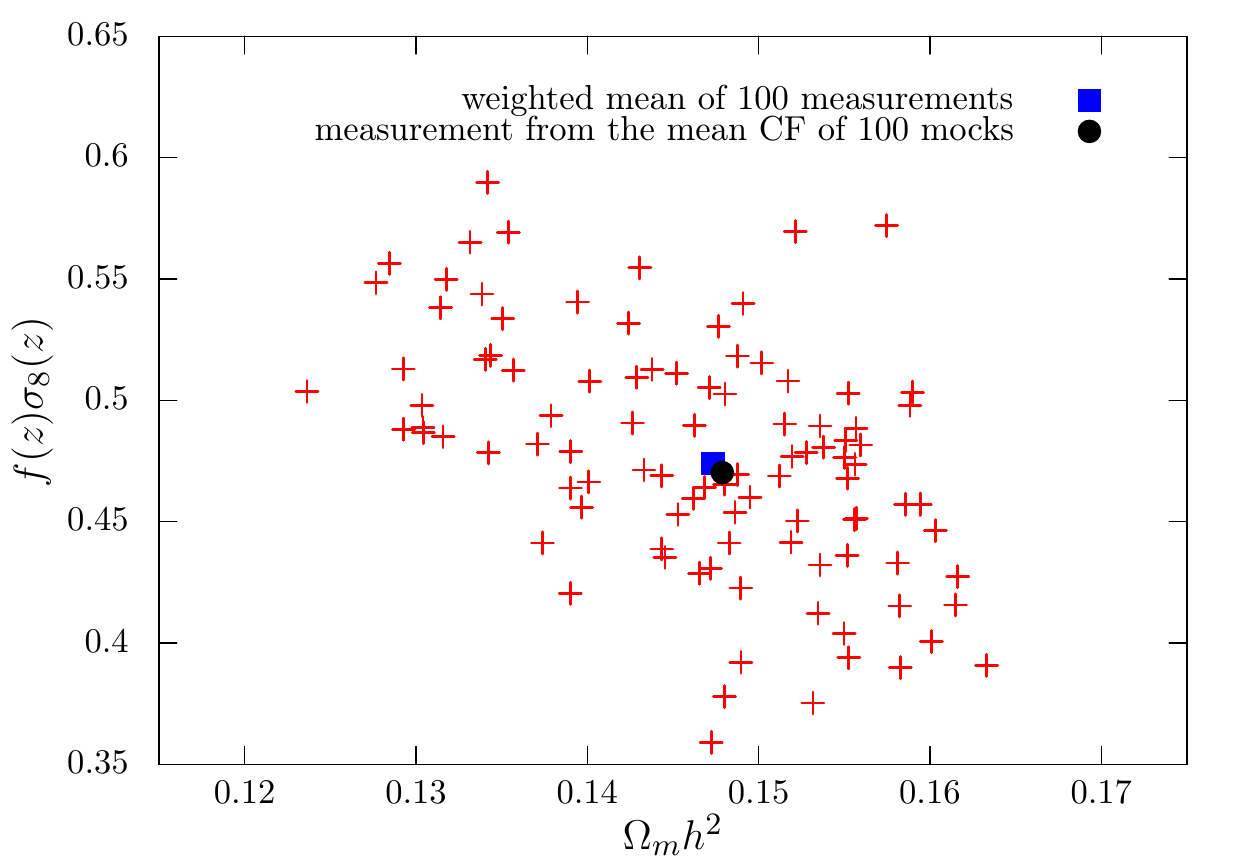}}
\caption{
Left panel: The small red crosses indicate the measurements of $\frac{H(z)r_s}{r_{s,fid}}$ and $\frac{D_A(z)r_{s,fid}}{r_s}$ from 100 individual CMASS mock catalogues. We show their weighted mean (with inverse variance weighting; blue square) and the measurement from the mean correlation function of the 100 mock catalogues (black circle); 
Right panel: The small red crosses indicate the measurements of  $\Omega_mh^2$ and f$\sigma_8$ from 100 individual CMASS mock catalogues. 
}
\label{fig:mock_100}
\end{figure*}

In this section, we will test our methodology by applying it to the mock catalogues. 
We first demonstrate that using the mean of the correlation functions is equivalent to using individual correlation functions from the mocks. 
We obtain the measurements from the first 100 CMASS mock catalogues within $0.43<z<0.75$. We use the fast model only and do not include the polynomial modelling of the systematics in these tests.
The left panel of Fig.~\ref{fig:mock_100} shows the distribution of the measurements of $H(z_{eff}) r_s/r_{s,fid}$ and $D_A(z_{eff}) r_{s,fid}/r_s$, where $r_s$ is the comoving sound horizon at the drag epoch and $r_{s,fid}=147.66$ Mpc is the sound scale of the fiducial cosmology used in this 
study. We also show the measurements from the weighted mean (using inverse variance weighting) of 100 correlation functions from these mocks. One can see that the weighted mean of the 100 individual measurements (blue square) is very close to the measurement from the mean correlation function from 100 mocks (black circle). We conclude that one can use the mean correlation function to represent the tests for multiple correlation functions.
The right panel of Fig.~\ref{fig:mock_100} shows the scatter of the measurements of $\Omega_mh^2$ and $f\sigma_8(z)$ from the same analysis above.

Note that the computing time is still expensive even after speeding up the analysis using the fast-slow model method described in previous sections.
Therefore, instead of applying the test using the correlation function from an individual mock catalogue, we use the mean of  the correlation functions from all the mocks. From these tests, we can see whether our methodology would introduce some bias or not. 
A small bias can be better detected using 2000 rather than 100 mock correlation functions. Therefore, 
we validate our methodology by applying our methodology on the mean correlation functions from 2000 mocks for different redshift bins and present the results in Table~\ref{table:2000mock}. One can see that for all the parameters in all the redshift bins, we recover the input parameters to within $0.3\sigma$. We show the results using the calibrated dewiggle model in Appendix~\ref{sec:table_of_dw} which also recovers the input parameters within reasonable precision, $0.6\sigma$. However, given that they are more realistic, we use the results from the Gaussian streaming model as our fiducial results.

\begin{table*}
 \begin{center}
  \begin{tabular}{c c c c c c} 
     \hline
	&	$\Omega_mh^2$		&		$f\sigma_8(z)$		&		$\frac{H(z)r_s}{r_{s,fid}}$		&		$\frac{D_A(z)r_{s,fid}}{r_s}$		&		$\frac{D_V(z)r_{s,fid}}{r_s}$			\\ \hline
$0.15<z<0.43$	&$	0.143	\pm	0.016	$&$	0.465	\pm	0.085	$&$	80.6	\pm	4.7	$&$	989	\pm	31	$&$	1267	\pm	29	$\\
input values	&$	0.14105			$&$	0.481			$&$	80.16			$&$	990.2			$&$	1269.19			$\\
deviation $\&$ uncertainty ($\%$)	&$	1.2	$ \& $	11.5	$&$	-3.3	$ \& $	17.6	$&$	0.5	$ \& $	5.8	$&$	-0.1	$ \& $	3.2	$&$	-0.2	$ \& $	2.3	$\\ \hline
$0.43<z<0.75$	&$	0.139	\pm	0.013	$&$	0.478	\pm	0.061	$&$	94.2	\pm	3.4	$&$	1416	\pm	25	$&$	2119	\pm	30	$\\
input values	&$	0.14105			$&$	0.4786			$&$	94.09			$&$	1409.26			$&$	2113.37			$\\
deviation $\&$ uncertainty ($\%$)	&$	-1.2	$ \& $	9.0	$&$	-0.2	$ \& $	12.8	$&$	0.1	$ \& $	3.7	$&$	0.5	$ \& $	1.7	$&$	0.3	$ \& $	1.4	$\\ \hline
$0.15<z<0.30$	&$	0.139	\pm	0.016	$&$	0.460	\pm	0.105	$&$	80.0	\pm	10.7	$&$	792	\pm	69	$&$	957	\pm	76	$\\
input values	&$	0.14105			$&$	0.4751			$&$	76.63			$&$	807.25			$&$	979.874			$\\
deviation $\&$ uncertainty ($\%$)	&$	-1.5	$ \& $	11.3	$&$	-3.3	$ \& $	22.1	$&$	4.4	$ \& $	14.0	$&$	-1.9	$ \& $	8.6	$&$	-2.3	$ \& $	7.8	$\\ \hline
$0.30<z<0.43$	&$	0.142	\pm	0.015	$&$	0.493	\pm	0.111	$&$	83.6	\pm	7.9	$&$	1090	\pm	49	$&$	1438	\pm	57	$\\
input values	&$	0.14105			$&$	0.4829			$&$	82.52			$&$	1088.59			$&$	1440.62			$\\
deviation $\&$ uncertainty ($\%$)	&$	1.0	$ \& $	10.9	$&$	2.1	$ \& $	22.9	$&$	1.3	$ \& $	9.5	$&$	0.2	$ \& $	4.5	$&$	-0.2	$ \& $	4.0	$\\ \hline
$0.43<z<0.55$	&$	0.140	\pm	0.016	$&$	0.478	\pm	0.084	$&$	88.1	\pm	4.9	$&$	1286	\pm	39	$&$	1830	\pm	41	$\\
input values	&$	0.14105			$&$	0.4827			$&$	88.59			$&$	1283.41			$&$	1823.53			$\\
deviation $\&$ uncertainty ($\%$)	&$	-0.7	$ \& $	11.3	$&$	-1.0	$ \& $	17.4	$&$	-0.5	$ \& $	5.6	$&$	0.2	$ \& $	3.1	$&$	0.4	$ \& $	2.3	$\\ \hline
$0.55<z<0.75$	&$	0.136	\pm	0.015	$&$	0.490	\pm	0.078	$&$	98.5	\pm	5.8	$&$	1462	\pm	42	$&$	2238	\pm	43	$\\
input values	&$	0.14105			$&$	0.4754			$&$	96.97			$&$	1461.99			$&$	2248.92			$\\
deviation $\&$ uncertainty ($\%$)	&$	-0.7	$ \& $	10.4	$&$	1.7	$ \& $	16.4	$&$	1.6	$ \& $	6.0	$&$	0.0	$ \& $	2.9	$&$	-0.5	$ \& $	1.9	$\\ \hline
     \hline
 \end{tabular}
 \end{center}
\caption{Measurements of $\Omega_mh^2$, $f(z)\sigma_8(z)$, $\frac{H(z)r_s}{r_{s,fid}}$, $\frac{D_A(z)r_{s,fid}}{r_s}$, and $\frac{D_V(z)r_{s,fid}}{r_s}$ from the mean of 2000 correlation functions, where the unit of $H(z)$ is $\Hunit$ and the units of $D_A(z)$ and $D_V(z)$ are Mpc. The effective redshifts are $\{$0.32, 0.59, 0.24, 0.37, 0.49, 0.64$\}$.
We show the means and standard deviations, input values, the differences between mean and input values (in percentage), and the standard deviations in percentage. 
}
\label{table:2000mock}
  \end{table*}

\subsection{Measurements of Cosmological Parameters from BOSS galaxy clustering}
\label{sec:results_hda}

We now present the dark energy model independent measurements of the parameters
$\{H(z)$, $D_A(z)$, $\Omega_m h^2$, $\beta(z)$, and $b\sigma_8(z)\}$, obtained by using the 
method described in previous sections. We also present derived 
parameters including $H(z)\frac{r_s}{r_{s,fid}}$, $D_A(z)\frac{r_{s,fid}}{r_{s}}$, $f(z)\sigma_8(z)$, and 
$D_V(z)\frac{r_{s,fid}}{r_{s}}$ with
\begin{equation} \label{eq:dv}
 D_V(z)\equiv \left[(1+z)^2D_A(z)^2\frac{cz}{H(z)}\right]^\frac{1}{3},
\end{equation} 
where $r_s$ is the comoving sound horizon at the drag epoch calculated by CAMB
and $r_{s,fid}=147.66$Mpc is the $r_s$ of the fiducial cosmology used in this study (same as the one used by the mock catalogues). We use $r_s/r_{s,fid}$ instead of $r_s$ since it is more insensitive to the approximate formula used for computing $r_s$.
$D_V(z)$ is the effective distance which can be measured from the spherical averaged correlation function or power spectrum (e.g. see \citealt{Eisenstein:2005su}).

Table \ref{table:results} lists the mean and standard deviation obtained from the MCMC likelihood analysis from the DR12 galaxy correlation function.
We measure $\{D_A(z)r_{s,fid}/r_s$, $H(z)r_s/r_{s,fid}$, $f(z)\sigma_8(z)$, $\Omega_m h^2\}$, $D_V r_s^{fid}/r_s$, $ \beta $, $b\sigma_8$ (they are not independent), using the range $40h^{-1}$Mpc $<s<180h^{-1}$Mpc, 
at the different redshift bins, i.e. $0.15<z<0.43$, $0.43<z<0.75$, $0.15<z<0.30$, $0.30<z<0.43$, $0.43<z<0.55$, $0.55<z<0.75$. The effective redshifts are $\{$0.32, 0.59, 0.24, 0.37, 0.49, 0.64$\}$.
The covariance matrices for these measurements can be found in the Appendix.

To conveniently compare with other measurements using CMASS sample within $0.43<z<0.7$ (we are using $0.43<z<0.75$), we extrapolated our measurements at $z=0.57$: $H(0.57)r_s/r_{s,fid}= 95.5 \pm 2.7\Hunit$, $D_A(0.57)r_{s,fid}/r_s=1404\pm23$Mpc, and $D_V(0.57)r_{s,fid}/r_s=2050\pm22$Mpc (see Table 9 of \citealt{Acacia}).

In the next section, we will describe how to use our results of single-probe measurements combining with other data set to constrain the parameters of given dark energy models.

 \begin{table*}
 \begin{center}
  \begin{tabular}{c | c c | c c c c} 
     \hline
	&$	0.15<z<0.43			$&$	0.43<z<0.75			$&$	0.15<z<0.30			$&$	0.30<z<0.43			$&$	0.43<z<0.55			$&$	0.55<z<0.75			$\\	\hline
$D_A r_s^{fid}/r_s$ 	&$	956	\pm	28	$&$	1421	\pm	23	$&$	826	\pm	45	$&$	993	\pm	65	$&$	1288	\pm	31	$&$	1444	\pm	41	$\\	
$H r_s/r_s^{fid}$ 	&$	75.0	\pm	4.0	$&$	96.7	\pm	2.7	$&$	78.8	\pm	5.6	$&$	74.8	\pm	6.3	$&$	87.5	\pm	4.8	$&$	98.4	\pm	3.7	$\\	
$f\sigma_8$ 	&$	0.397	\pm	0.073	$&$	0.497	\pm	0.058	$&$	0.493	\pm	0.105	$&$	0.378	\pm	0.076	$&$	0.456	\pm	0.068	$&$	0.454	\pm	0.064	$\\	
$\Omega_m h^2$ 	&$	0.143	\pm	0.017	$&$	0.137	\pm	0.015	$&$	0.136	\pm	0.017	$&$	0.147	\pm	0.014	$&$	0.144	\pm	0.016	$&$	0.140	\pm	0.017	$\\	
$D_V r_s^{fid}/r_s$ 	&$	1268	\pm	26	$&$	2106	\pm	23	$&$	987	\pm	40	$&$	1402	\pm	69	$&$	1837	\pm	36	$&$	2220	\pm	39	$\\	
$ \beta $ 	&$	0.301	\pm	0.066	$&$	0.435	\pm	0.070	$&$	0.389	\pm	0.096	$&$	0.287	\pm	0.067	$&$	0.367	\pm	0.072	$&$	0.410	\pm	0.077	$\\	
$b\sigma_8$ 	&$	1.332	\pm	0.099	$&$	1.154	\pm	0.090	$&$	1.287	\pm	0.129	$&$	1.332	\pm	0.137	$&$	1.256	\pm	0.112	$&$	1.120	\pm	0.094	$\\		
     \hline
 \end{tabular}
 \end{center}
\caption{Our measurements of $\{D_A(z)r_{s,fid}/r_s$, $H(z)r_s/r_{s,fid}$, $f(z)\sigma_8(z)$, $\Omega_m h^2\}$, $D_V r_{s,fid}/r_s$, $ \beta $, $b\sigma_8$, from BOSS DR12 data at the different redshift bins stated, using the range $40h^{-1}$Mpc $<s<180h^{-1}$Mpc; $r_{s,fid}$ is 147.66 Mpc in this study; the unit of $H(z)$ is $\Hunit$ and the units of $D_A(z)$ and $D_V(z)$ are Mpc. The effective redshifts of these redshift bins are $\{$0.32, 0.59, 0.24, 0.37, 0.49, 0.64$\}$.}
\label{table:results}
  \end{table*}
 
\section{Constraining cosmological parameters of given dark energy models}
\label{sec:application}
\subsection{Likelihood derivation}
In this section, we describe the steps to combine our results with other data sets assuming some dark energy models. 
Here, we use the results from two redshift bins, $0.15<z<0.43$ (LOWZ) and $0.43<z<0.75$ (CMASS), as an example.
For a given model and cosmological parameters, one can compute 
$H(z)\frac{r_s}{r_{s,fid}}$, $D_A(z)\frac{r_{s,fid}}{r_{s}}$, $f(z)\sigma_8(z)$, and $\Omega_mh^2$. From Tables \ref{table:cov_z15z43} and \ref{table:cov_z43z75} in Appendix \ref{sec:norm_covar} and the standard deviations in Table \ref{table:results},
one can compute the covariance matrices, $M_{ij,\textrm{CMASS}}$ and $M_{ij,\textrm{LOWZ}}$, of these four parameters. Then, $\chi^2_{\textrm{CMASS}}$ and $\chi^2_{\textrm{LOWZ}}$ can be computed by
\begin{equation}
 \chi^2_{\textrm{CMASS}}=\Delta_{\textrm{CMASS}}M_{ij,\textrm{CMASS}}^{-1}\Delta_{\textrm{CMASS}},
\end{equation}
and
\begin{equation}
 \chi^2_{\textrm{LOWZ}}=\Delta_{\textrm{LOWZ}}M_{ij,\textrm{LOWZ}}^{-1}\Delta_{\textrm{LOWZ}},
\end{equation}
where 
\begingroup
\everymath{\scriptstyle}
\small
\begin{equation}
 \Delta_{\textrm{CMASS}}=
\left(\begin{array}{c}
D_A(z)\frac{r_{s,fid}}{r_{s}}-1421 \\ 
H(z)\frac{r_s}{r_{s,fid}}-96.7 \\ 
f(z)\sigma_8(z)-0.497\\
\Omega_mh^2-0.137
\end{array}\right),
\end{equation}
\endgroup
\begingroup
\everymath{\scriptstyle}
\small
\begin{equation}
 \Delta_{\textrm{LOWZ}}=
\left(\begin{array}{c}
D_A(z)\frac{r_{s,fid}}{r_{s}}-956 \\ 
H(z)\frac{r_s}{r_{s,fid}}-75.0 \\ 
f(z)\sigma_8(z)-0.397\\
\Omega_mh^2-0.143
\end{array}\right),
\end{equation}
\endgroup
\begingroup
\everymath{\scriptstyle}
\small
\begin{eqnarray}
M_{ij,\textrm{CMASS}}=\nonumber \\
\left(\begin{array}{cccc}
    0.53559E+03 &   0.27875E+02 &   0.70092E+00 &  -0.29507E-01    \nonumber \\
    0.27875E+02 &   0.74866E+01 &   0.85855E-01 &  -0.92898E-02   \\
    0.70092E+00 &   0.85855E-01 &   0.33643E-02 &  -0.51341E-03    \\
   -0.29507E-01 &  -0.92898E-02 &  -0.51341E-03 &   0.22673E-03
\end{array}\right),
\end{eqnarray}
\endgroup
and
\begingroup
\everymath{\scriptstyle}
\small
\begin{eqnarray}
M_{ij,\textrm{LOWZ}}=\nonumber \\
\left(\begin{array}{cccc}
    0.77636E+03 &   0.43792E+02 &   0.11413E+01 &   0.86090E-01 \nonumber \\   
    0.43792E+02 &   0.16253E+02 &   0.19856E+00 &   0.21477E-01   \\
    0.11413E+01 &   0.19856E+00 &   0.53875E-02 &   0.69008E-04   \\
    0.86090E-01 &   0.21477E-01 &   0.69008E-04  &  0.29001E-03
\end{array}\right).
\end{eqnarray}
\endgroup

One can include the cosmological constraints from the SDSS/BOSS galaxy clustering by adding $\chi^2_{\textrm{LOWZ}}+\chi^2_{\textrm{CMASS}}$ in the MCMC analysis, due to the negligible correlation of these samples.

\subsection{Constraining Dark Energy Parameters combining with external data sets} 
\label{sec:models}
In this section, we present examples of combining our galaxy clustering results with the Planck CMB data assuming specific dark energy models. 
The Planck data set we use is the \textit{Planck} 2015 measurements \citep{Adam:2015rua, Ade:2015xua}.
The reference likelihood code \citep{Aghanim:2015xee} was downloaded from the
Planck Legacy Archive\footnote{PLA: \url{http://pla.esac.esa.int/}}.
Here we combine the \textit{Plik} baseline likelihood for high multipoles ($30 \le \ell \le 2500$)
using the TT, TE and EE power spectra, and the \textit{Planck} low-$\ell$ multipole likelihood in the
range $2 \le \ell \le 29$ (hereafter lowTEB). We also include the \textit{Planck} 2015 lensing
likelihood \citep{Ade:2015zua}, constructed from the measurements of the power spectrum of the
lensing potential (hereafter referred as "lensing"). When using the \textit{Planck} lensing likelihood, the
$A_{\rm lens}$ parameter is always set to 1 \citep{Ade:2015xua}.  

\begin{figure*}
\centering
\subfigure{\includegraphics[width=1 \columnwidth,clip,angle=0]{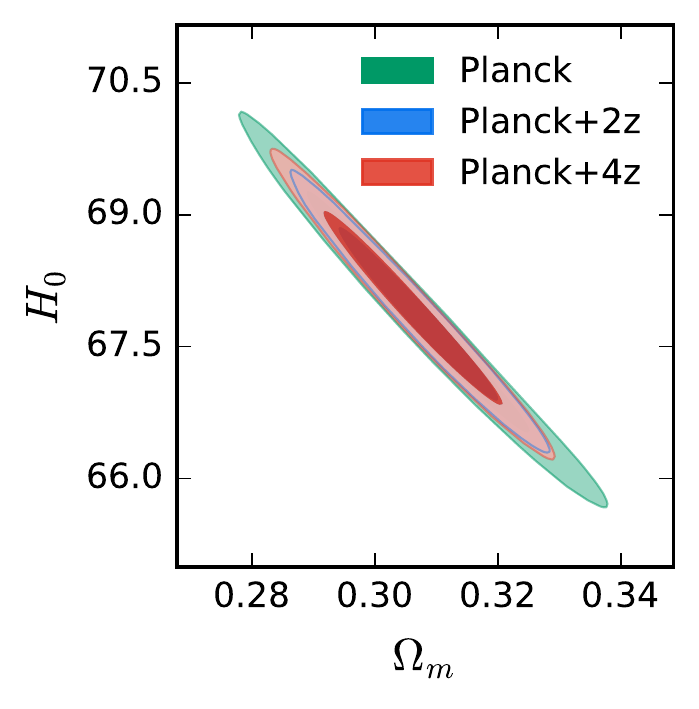}}
\subfigure{\includegraphics[width=1 \columnwidth,clip,angle=0]{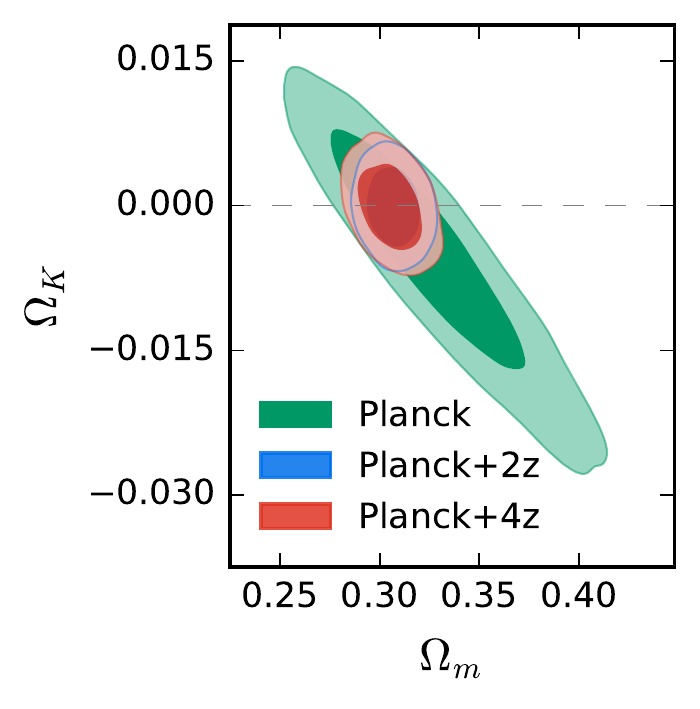}}
\caption{
Left panel: 2D marginalized
  contours for $68\%$ and $95\%$ confidence levels for $\Omega_m$ and $H_0$ ($\Lambda$CDM model assumed)
from Planck-only (green), Planck+CMASS (1 bin)+LOWZ (1 bin) (blue), and Planck+CMASS (2 bins)+LOWZ (2 bins) (red); right panel: 2D marginalized
  contours for $68\%$ and $95\%$ confidence level for $\Omega_m$ and $\Omega_k$ (o$\Lambda$CDM model assumed). One can see that $\Omega_k$ is consistent with 0 which corresponds to the flat universe.
}
\label{fig:lcdm_olcdm}
\end{figure*}

\begin{figure*}
\centering
\subfigure{\includegraphics[width=1 \columnwidth,clip,angle=0]{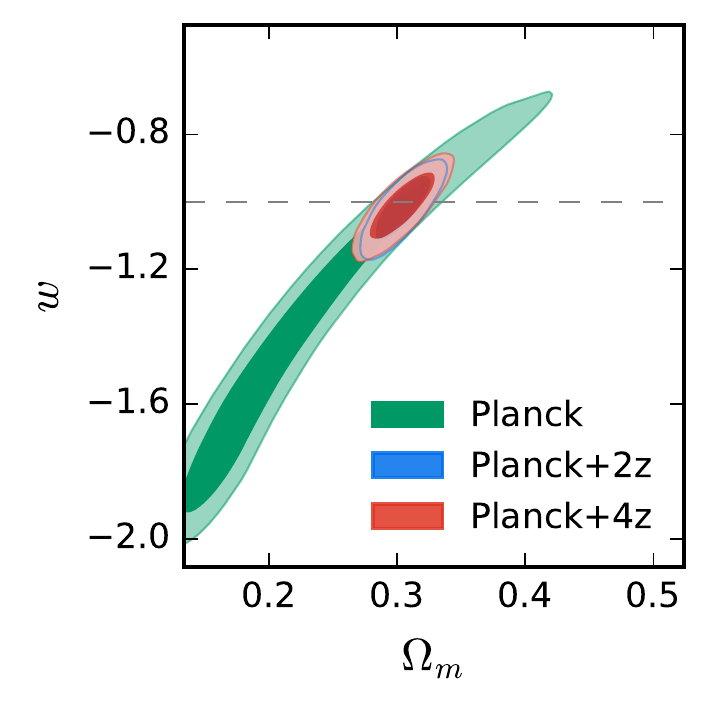}}
\subfigure{\includegraphics[width=1 \columnwidth,clip,angle=0]{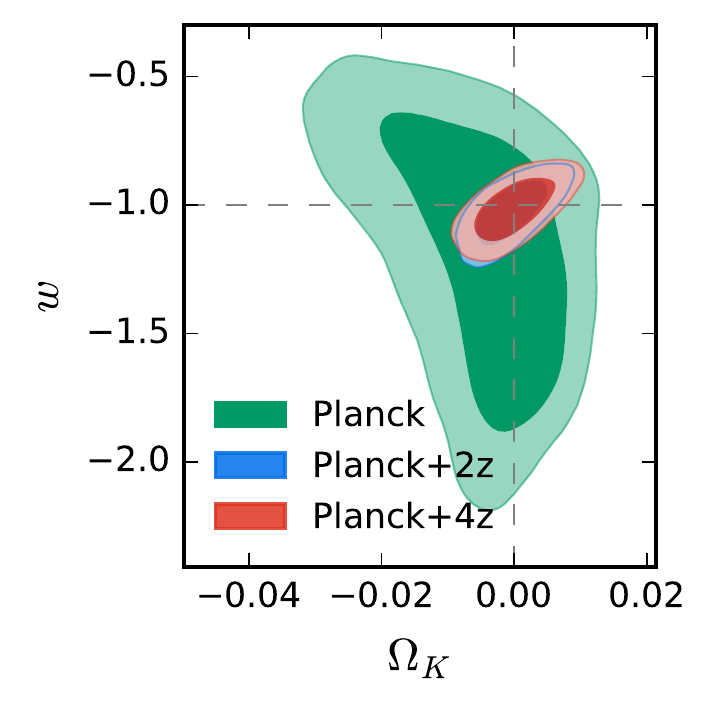}}
\caption{
Left panel: 2D marginalized
  contours for $68\%$ and $95\%$ confidence level for $\Omega_m$ and $w$ ($w$CDM model assumed)
from Planck-only (green), Planck+CMASS (1 bin)+LOWZ (1 bin) (blue), and Planck+CMASS (2 bins)+LOWZ (2 bins) (red); right panel: 2D marginalized
  contours for $68\%$ and $95\%$ confidence level for $\Omega_k$ and $w$ (o$w$CDM model assumed). One can see that $\Omega_k$ is consistent with 0 and $w$ is consistent with -1 which corresponds to the $\Lambda$CDM.
}
\label{fig:wcdm_owcdm}
\end{figure*}

\begin{figure*}
\centering
\subfigure{\includegraphics[width=1 \columnwidth,clip,angle=0]{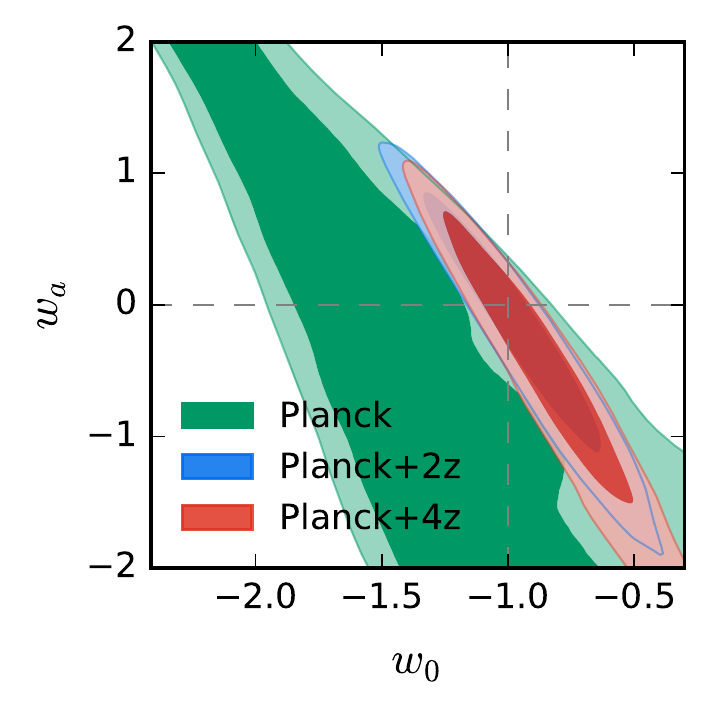}}
\subfigure{\includegraphics[width=1 \columnwidth,clip,angle=0]{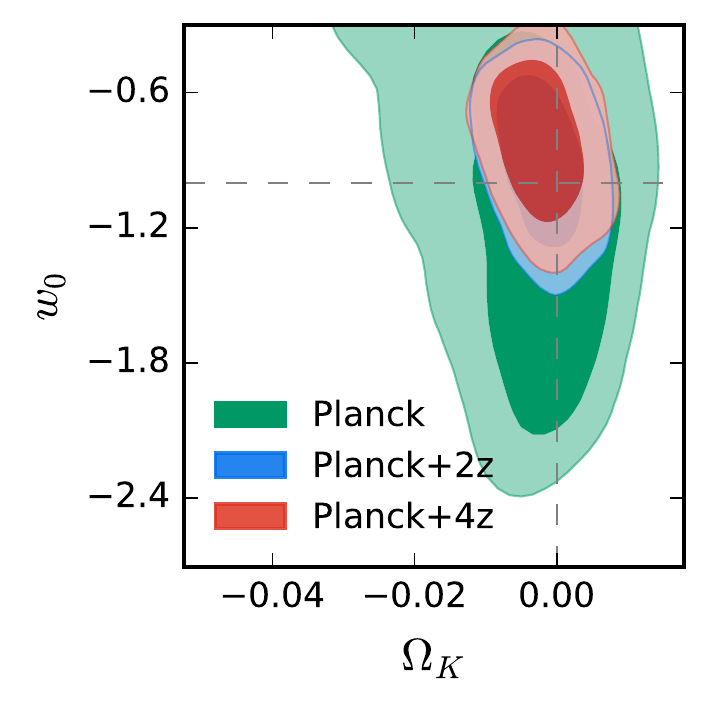}}
\caption{
Left panel: 2D marginalized
  contours for $68\%$ and $95\%$ confidence level for $w_0$ and $w_a$ ($w_0w_a$CDM model assumed)
from Planck-only (green), Planck+CMASS (1 bin)+LOWZ (1 bin) (blue), and Planck+CMASS (2 bins)+LOWZ (2 bins) (red); right panel: 2D marginalized
  contours for $68\%$ and $95\%$ confidence level for $\Omega_k$ and $w_0$ (o$w_0w_a$CDM model assumed).
One can see that $w_0$ and $w_a$ are consistent with -1 and 0 respectively which correspond to the $\Lambda$CDM.
}
\label{fig:w0wacdm_ow0wacdm}
\end{figure*}

Table \ref{table:DEmodels} shows the cosmological constraints assuming 
flat $\Lambda$CDM, o$\Lambda$CDM (non-flat $\Lambda$CDM), $w$CDM (constant equation of state of dark energy), o$w$CDM (non-flat $w$CDM), $w_0w_a$CDM ( time-dependent equation
of state) and o$w_0w_a$CDM (non-flat $w_0w_a$CDM). In addition to using 2 redshift bins, we use 4 redshift bins but we do not find any improvement in terms of constraining cosmological parameters. 
It should indicate that the the models we are testing are still simple and do not benefit from higher redshift sensitivity. In addition, some information (pair counts) would be lost when we slice the sample into more bins.
In Fig.~\ref{fig:lcdm_olcdm}, \ref{fig:wcdm_owcdm}, and \ref{fig:w0wacdm_ow0wacdm}, we show 2D marginalized
  contours for $68\%$ and $95\%$ confidence levels for $\Omega_m$ and $H_0$ ($\Lambda$CDM model assumed); $\Omega_m$ and $\Omega_k$ (o$\Lambda$CDM model assumed); $\Omega_m$ and $w$ ($w$CDM model assumed); $\Omega_k$ and $w$ (o$w$CDM model assumed); $w_0$ and $w_a$ ($w_0w_a$CDM model assumed); $\Omega_k$ and $w_0$ (o$w_0w_a$CDM model assumed). One can see that all the constraints are consistent with  flat $\Lambda$CDM.

\begin{table*}
\begin{center}
\begin{tabular}{lrrrrrr}
\hline
		&	$\Omega_m$	&			$H_0$	&			$\sigma_8$	&			$\Omega_k$	&			$w$ or $w_0$	&			$w_a$			\\	\hline
Planck+2bins	($\Lambda$CDM)	& $	0.307	\pm	0.008	$&$	67.9	\pm	0.6	$&$	0.815	\pm	0.009	$&$	0			$&$	-1			$&$	0			$\\	
Planck+4bins	($\Lambda$CDM)	& $	0.306	\pm	0.009	$&$	67.9	\pm	0.7	$&$	0.815	\pm	0.009	$&$	0			$&$	-1			$&$	0			$\\	
Planck+2bins	(o$\Lambda$CDM)	& $	0.307	\pm	0.008	$&$	67.8	\pm	0.8	$&$	0.815	\pm	0.009	$&$	0.000	\pm	0.003	$&$	-1			$&$	0			$\\	
Planck+4bins	(o$\Lambda$CDM)	& $	0.306	\pm	0.010	$&$	68.0	\pm	1.0	$&$	0.815	\pm	0.010	$&$	0.000	\pm	0.003	$&$	-1			$&$	0			$\\	
Planck+2bins	($w$CDM)	& $	0.304	\pm	0.013	$&$	68.3	\pm	1.5	$&$	0.819	\pm	0.015	$&$	0			$&$	-1.02	\pm	0.06	$&$	0			$\\	
Planck+4bins	($w$CDM)	& $	0.304	\pm	0.016	$&$	68.3	\pm	1.7	$&$	0.818	\pm	0.017	$&$	0			$&$	-1.01	\pm	0.06	$&$	0			$\\
Planck+2bins	(o$w$CDM)	& $	0.305	\pm	0.015	$&$	68.2	\pm	1.5	$&$	0.819	\pm	0.017	$&$	0.000	\pm	0.003	$&$	-1.02	\pm	0.08	$&$	0			$\\	
Planck+4bins	(o$w$CDM)	& $	0.304	\pm	0.017	$&$	68.2	\pm	1.8	$&$	0.817	\pm	0.017	$&$	0.000	\pm	0.004	$&$	-1.02	\pm	0.08	$&$	0			$\\	
Planck+2bins	($w_0w_a$CDM)	& $	0.310	\pm	0.021	$&$	67.8	\pm	2.2	$&$	0.815	\pm	0.019	$&$	0			$&$	-0.95	\pm	0.22	$&$	-0.22	\pm	0.63	$\\	
Planck+4bins	($w_0w_a$CDM)	& $	0.314	\pm	0.021	$&$	67.2	\pm	2.2	$&$	0.810	\pm	0.019	$&$	0			$&$	-0.86	\pm	0.22	$&$	-0.50	\pm	0.67	$\\	
Planck+2bins	(o$w_0w_a$CDM)	& $	0.312	\pm	0.020	$&$	67.4	\pm	2.2	$&$	0.813	\pm	0.018	$&$	-0.002	\pm	0.004	$&$	-0.90	\pm	0.23	$&$	-0.49	\pm	0.75	$\\		
Planck+4bins	(o$w_0w_a$CDM)	& $	0.316	\pm	0.022	$&$	66.9	\pm	2.3	$&$	0.809	\pm	0.019	$&$	-0.002	\pm	0.004	$&$	-0.82	\pm	0.22	$&$	-0.73	\pm	0.73	$\\		
\hline
\end{tabular}
\end{center}
\caption{ 
The cosmological constraints from 2 redshift bins and 4 redshift bins combined with Planck data assuming $\Lambda$CDM, non-flat $\Lambda$CDM (o$\Lambda$CDM), $w$CDM, $w_0w_a$CDM, and o$w_0w_a$CDM.
The units of $H_0$ are $\Hunit$.
} \label{table:DEmodels}
\end{table*}

\section{COMPARISON WITH OTHER WORKS}
\label{sec:compare}
We compile the measurements of $f(z)\sigma_8(z)$, $D_A(z)/r_s$, $H(z)*r_s$, and $D_V(z)/r_s$ from various galaxy surveys in Tables \ref{table:fs8}, \ref{table:DVrs}, and \ref{table:H_DA} in the Appendix \ref{sec:compile_previous}. 
We have included the measurements from 
VIMOS-VLT Deep Survey (VVDS; \citealt{Guzzo:2008ac}),
2dFGRS \citep{Percival:2004fs},
Six-degree-Field Galaxy Survey (6dFGS; \citealt{Beutler:2011hx,Beutler:2012px}),
WiggleZ \citep{Blake:2011en,Blake:2011rj,Blake:2012pj,Contreras:2013bol},
SDSS-II/DR7 \citep{Percival:2009xn,Chuang:2010dv,Samushia:2011cs,Chuang:2011fy,Chuang:2012ad,Chuang:2012qt,Ross:2014qpa,Padmanabhan:2012hf,Xu:2012fw, 
Seo:2012xy,Hemantha:2013sea}
SDSS-III/BOSS \citep{Anderson:2012sa,Reid:2012sw, Anderson:2013oza, Chuang:2013hya, Kazin:2013rxa,Wang:2014qoa, Anderson:2013zyy,Beutler:2013yhm,Samushia:2013yga,Sanchez:2013tga,Tojeiro:2014eea,Reid:2014iaa,Alam:2015qta,Gil-Marin:2015nqa,Gil-Marin:2015sqa,Cuesta:2015mqa},
\cite{Acacia} (BOSS collaboration paper for final data release) and its companion papers including this paper and 
\cite{Ross16, 
Vargas-Magana16, 
Beutler16b, 
Satpathy16, 
Beutler16c, 
Sanchez16a, 
Grieb16, 
Pellejero-Ibanez16}.

\begin{figure*}
\centering
\includegraphics[width=1.7 \columnwidth,clip,angle=-0]{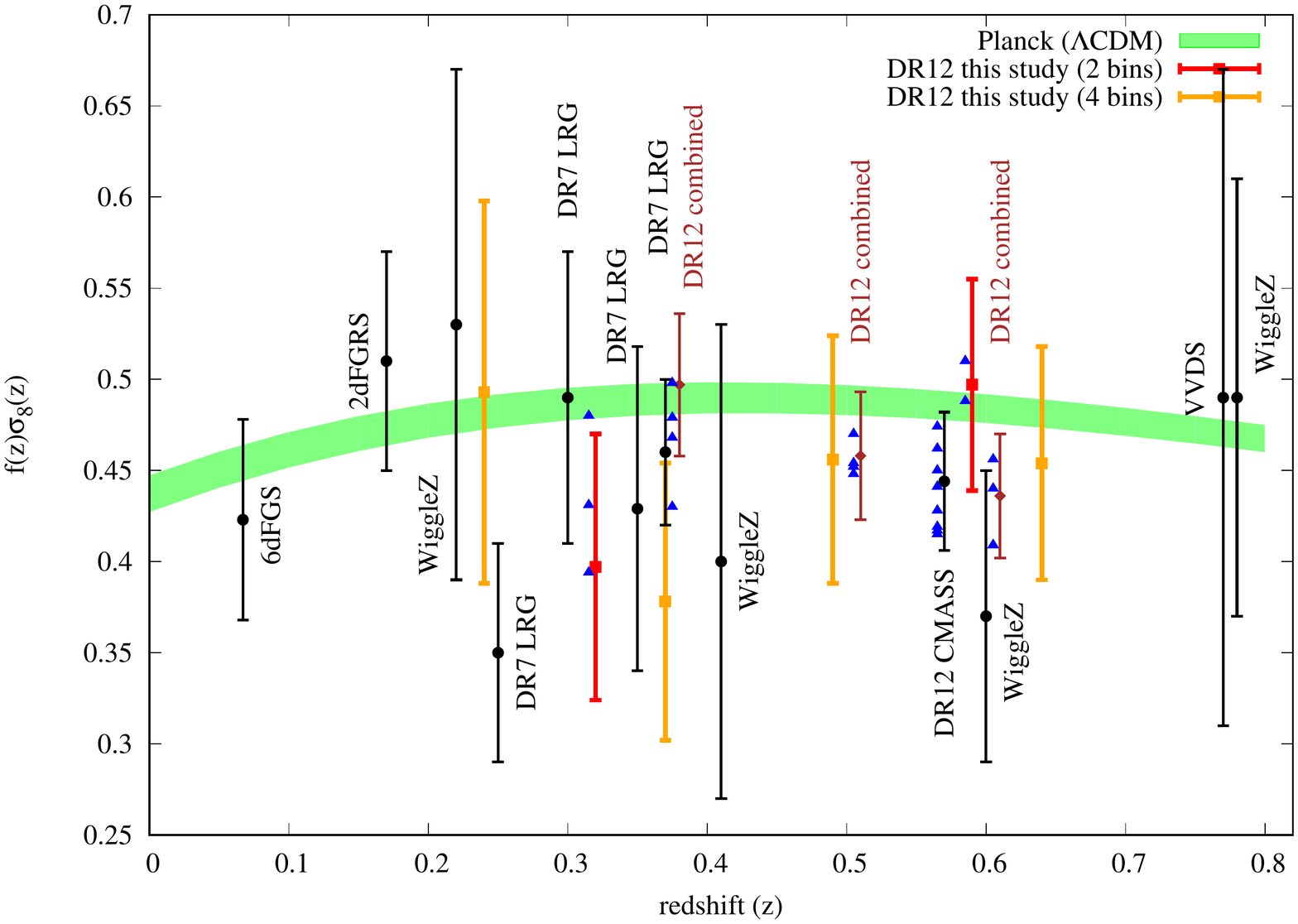}
\caption{
We compare the constraints of $f(z)\sigma_8(z)$ from CMB data (Planck) with our measurements (red squares), other measurements from SDSS galaxy sample, and the measurements compiled by \protect\cite{Samushia:2012iq} (black circles). We also compare with the consensus measurements from \protect\cite{Acacia} (BOSS collaboration paper for final data release; brown diamond points).
The constraints from CMB are obtained assuming a $\Lambda$CDM model.
}
\label{fig:compare_fs8_lcdm}
\end{figure*}

\begin{figure*}
\centering
\includegraphics[width=1.7 \columnwidth,clip,angle=-0]{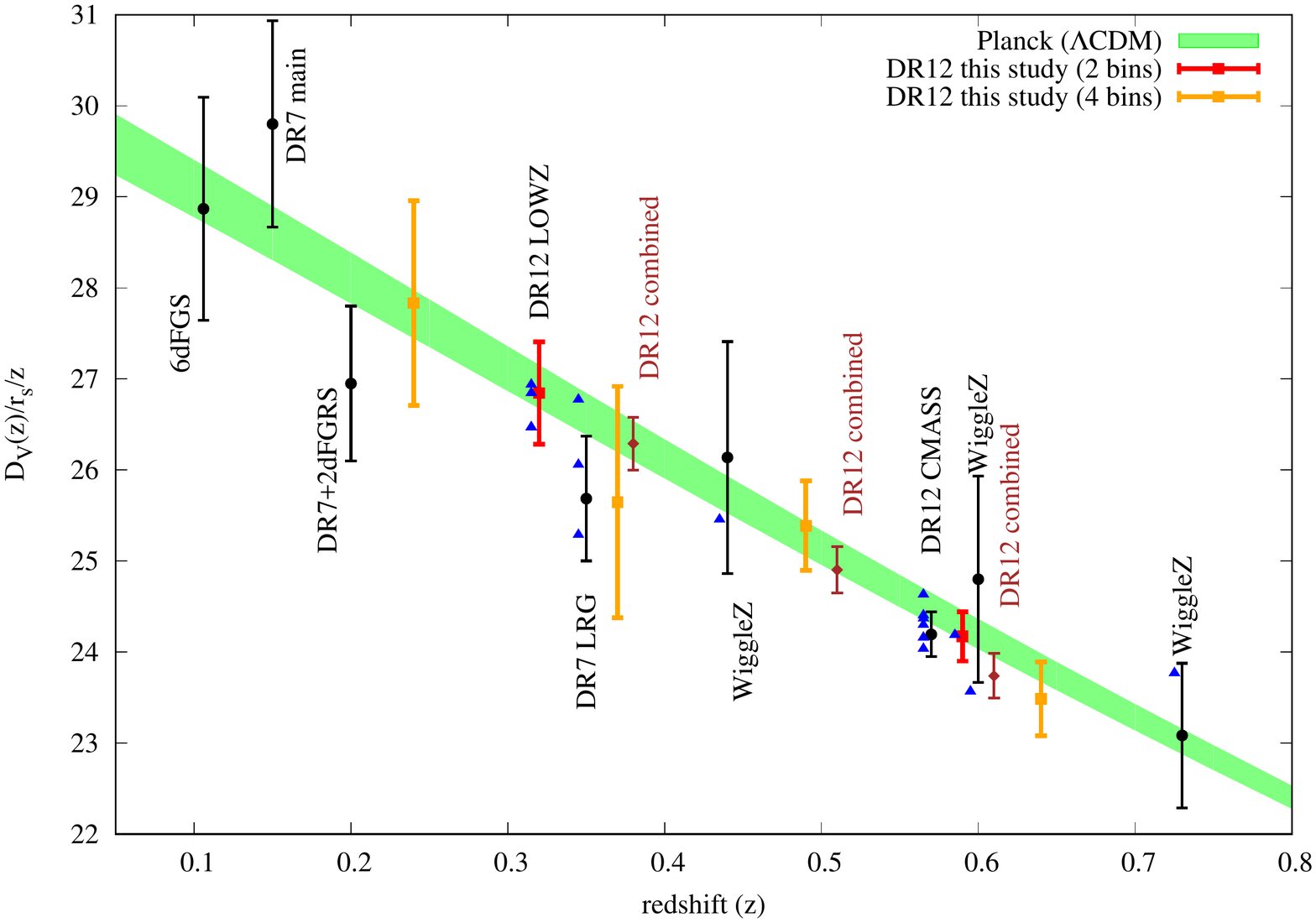}
\caption{
We compare the constraints of $\frac{D_V(z)}{r_s z}$ from CMB data (Planck) with our measurements (red squares), and other measurements (black circles and blue triangles; \citealt{Blake:2011en,Beutler:2011hx,Percival:2009xn,Chuang:2010dv,Chuang:2011fy,Padmanabhan:2012hf,Anderson:2012sa,Anderson:2013zyy,
Beutler:2013yhm,Samushia:2013yga,Tojeiro:2014eea,Ross:2014qpa}, 
and \protect\cite{Acacia} (BOSS collaboration paper for final data release).
).  The consensus values from \protect\cite{Acacia} are shown with brown diamond points.
When there are more than one measurement at the same redshift, we mark one of the measurement using a black circle with error bar (i.e., the measurement from \citealt{Chuang:2011fy} at $z=0.35$ and the consensus values from \citealt{Cuesta:2015mqa} at $z=0.57$) and mark the others with blue triangles with a slight shift in redshift to make the plot more clear.
The constraints from CMB are obtained assuming a $\Lambda$CDM model.
}
\label{fig:compare_dvrs_lcdm}
\end{figure*}

\begin{figure*}
\centering
\includegraphics[width=1.7 \columnwidth,clip,angle=-0]{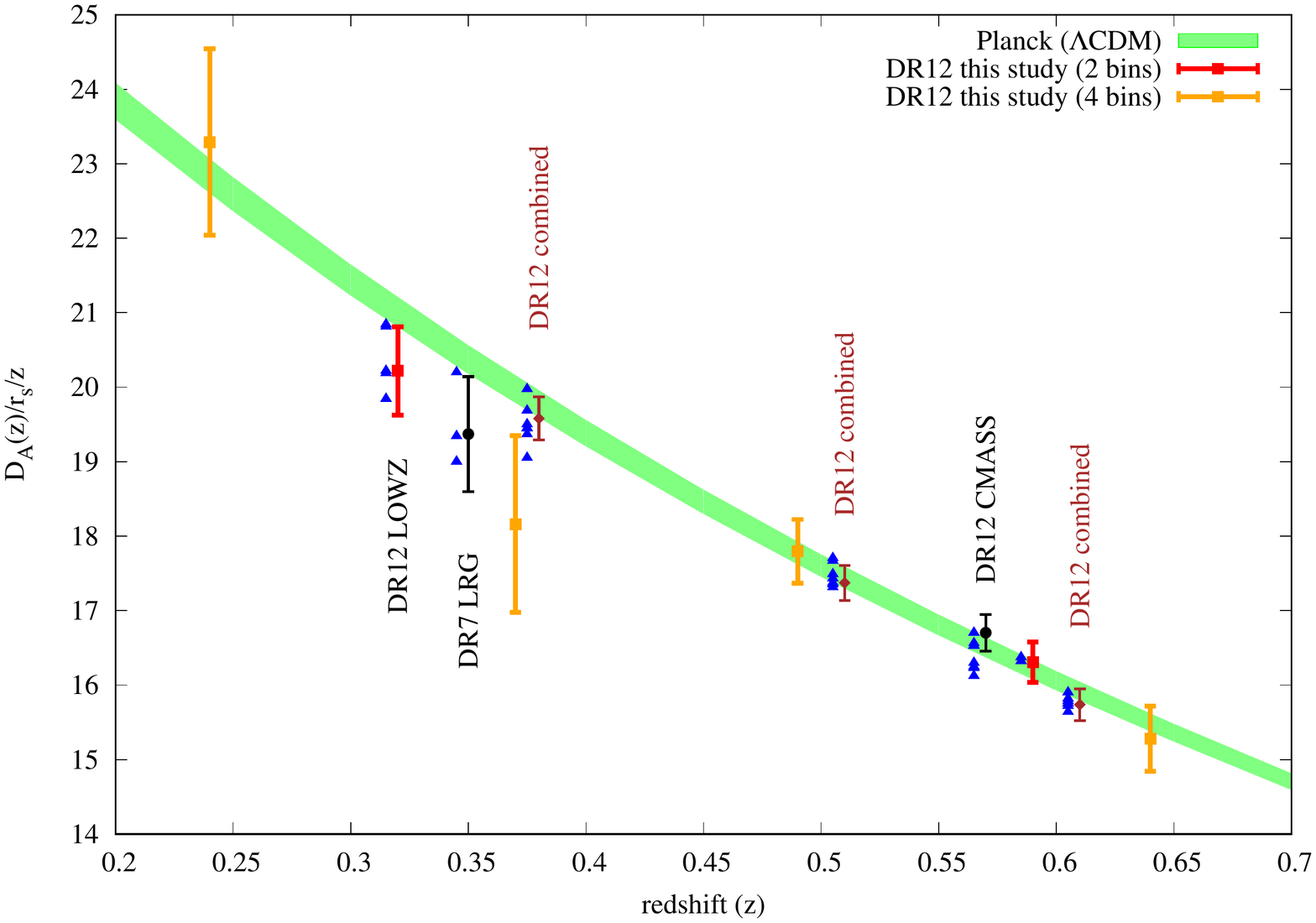}
\caption{
We compare the constraints of $\frac{D_A(z)}{r_s z}$ from CMB data (Planck) with our measurements (red squares), and other measurements (black circles and blue triangles; \citealt{Chuang:2011fy,Chuang:2012ad,Chuang:2012qt,Xu:2012fw,Hemantha:2013sea,Anderson:2013oza,Chuang:2013hya, Kazin:2013rxa,Wang:2014qoa, Anderson:2013zyy,Beutler:2013yhm,Gil-Marin:2015nqa,Gil-Marin:2015sqa,Cuesta:2015mqa},
\protect\cite{Acacia} (BOSS collaboration paper for final data release) and its companion papers including this paper and 
\protect\cite{Ross16, 
Vargas-Magana16, 
Beutler16b, 
Satpathy16, 
Beutler16c, 
Sanchez16a, 
Grieb16, 
Pellejero-Ibanez16}).
The consensus values from \protect\cite{Acacia} are shown with brown diamond points.
When there are more than one measurement at the same redshift, we mark one of the measurements using a black circle with error bar (i.e., the measurement from \citealt{Chuang:2011fy} at $z=0.35$ and the consensus values from \citealt{Cuesta:2015mqa} at $z=0.57$) and mark the others with blue triangles with a slight shift in redshift to make the plot more clear.
The constraints from CMB are obtained assuming a $\Lambda$CDM model.
}
\label{fig:compare_dars_lcdm}
\end{figure*}

\begin{figure*}
\centering
\includegraphics[width=1.7 \columnwidth,clip,angle=-0]{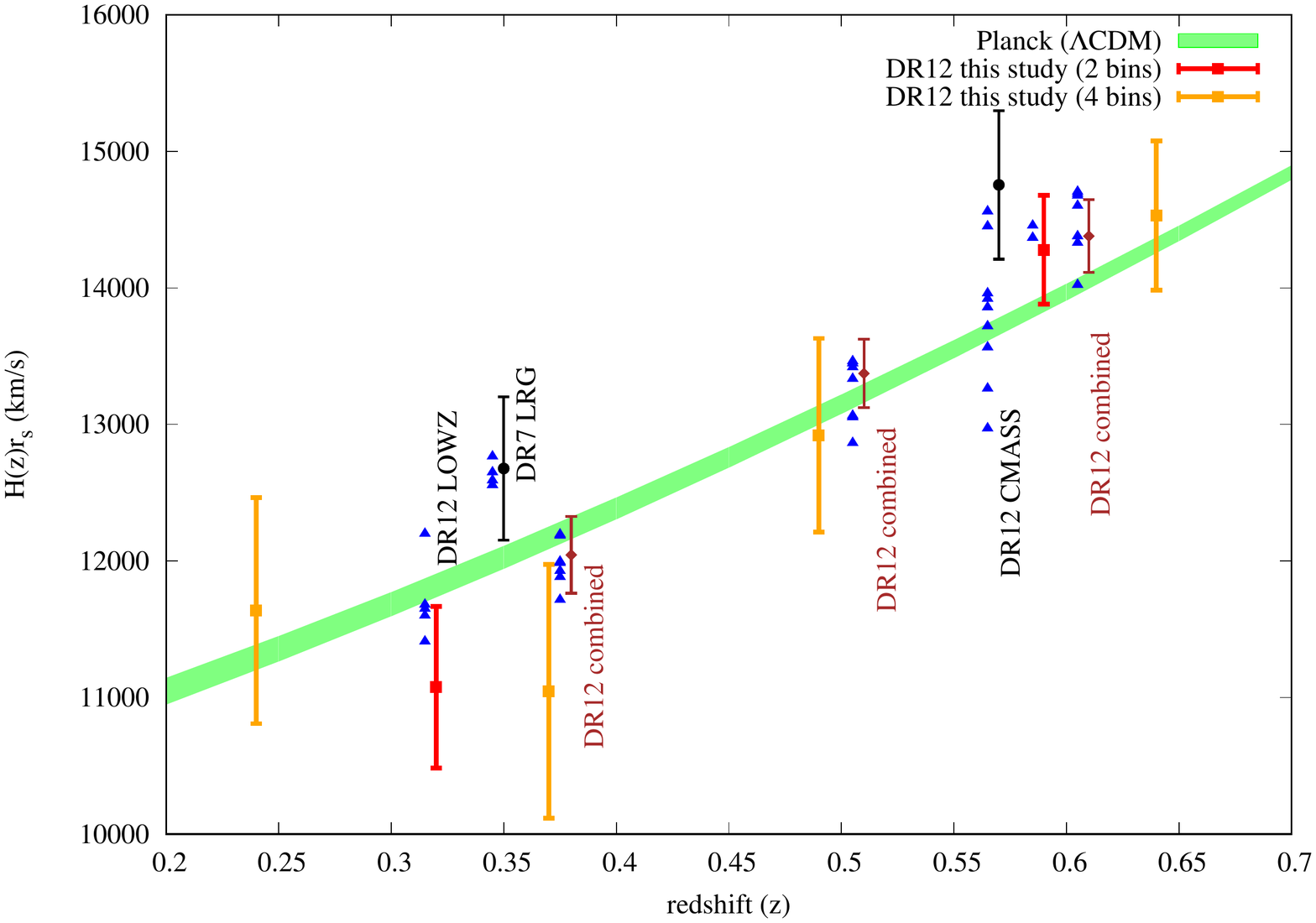}
\caption{
We compare the constraints of $H(z)r_s$ (km/s) from CMB data (Planck) with our measurements (red squares), and other measurements (black circles and blue triangles; \citealt{Chuang:2011fy,Chuang:2012ad,Chuang:2012qt,Xu:2012fw,Hemantha:2013sea,Anderson:2013oza,Chuang:2013hya, Kazin:2013rxa,Wang:2014qoa, Anderson:2013zyy,Beutler:2013yhm,Gil-Marin:2015nqa,Gil-Marin:2015sqa,Cuesta:2015mqa},
\protect\cite{Acacia} (BOSS collaboration paper for final data release) and its companion papers including this paper and 
\protect\cite{Ross16, 
Vargas-Magana16, 
Beutler16b, 
Satpathy16, 
Beutler16c, 
Sanchez16a, 
Grieb16, 
Pellejero-Ibanez16}).
The consensus values from \protect\cite{Acacia} are shown with brown diamond points.
When there are more than one measurement at the same redshift, we mark one of the measurements using a black circle with error bar (i.e., the measurement from \citealt{Chuang:2011fy} at $z=0.35$ and the consensus values from \citealt{Cuesta:2015mqa} at $z=0.57$) and mark the others with blue triangles with a slight shift in redshift to make the plot more clear.
The constraints from CMB are obtained assuming a $\Lambda$CDM model.
}
\label{fig:compare_hrs_lcdm}
\end{figure*}

To be able to include more measurements, we quote $D_V(z)/r_s$  instead of $D_V(z)r_{s,fid}/r_s$ since  $r_{s,fid}$ was not provided in some references.
In Figs. \ref{fig:compare_fs8_lcdm}, \ref{fig:compare_dvrs_lcdm}, \ref{fig:compare_dars_lcdm}, and \ref{fig:compare_hrs_lcdm},
we compare the constraints of $f(z)\sigma_8(z)$, $D_V(z)/r_s$, $D_A(z)/r_s$, and $H(z)r_s$ from CMB data (Planck assuming $\Lambda$CDM) with the measurements from galaxy clustering analyses compiled in Tables \ref{table:fs8}, \ref{table:DVrs}, and \ref{table:H_DA}.

In these figures, when there are many measurements that correspond to the same redshift, we show the mean and error bar for only one of them (as indicated in the caption) and show only the mean values indicated with triangles for the rest of the measurements. We also slightly shift the redshift to make the figures more clear.
Since we are using the CMASS galaxy sample with an extended redshift range ($0.43<z<0.75$) compared to other studies using the CMASS galaxy sample ($0.43<z<0.7$), the comparison cannot be done directly. However, our measurements agree very well with the prediction from Planck data assuming $\Lambda$CDM, and so do the measurements from previous works. 
One can see that the measurements of $D_V(z)/r_s$ from different analyses but at the same redshift agree with each other. However, the measurements of $H(z)r_s/r_{s,fid}$ and $D_A(z) r_{s,fid}/r_s$ have larger scatter. This is expected since $D_V(z)/r_s$ is driven by the BAO feature in the monopole. But, $H(z)r_s$ and $D_A(z)/r_s$ is correlated with the shape of BAO feature which has larger uncertainties among different models. 

\begin{figure*}
\centering
\subfigure{\includegraphics[width=1 \columnwidth,clip,angle=0]{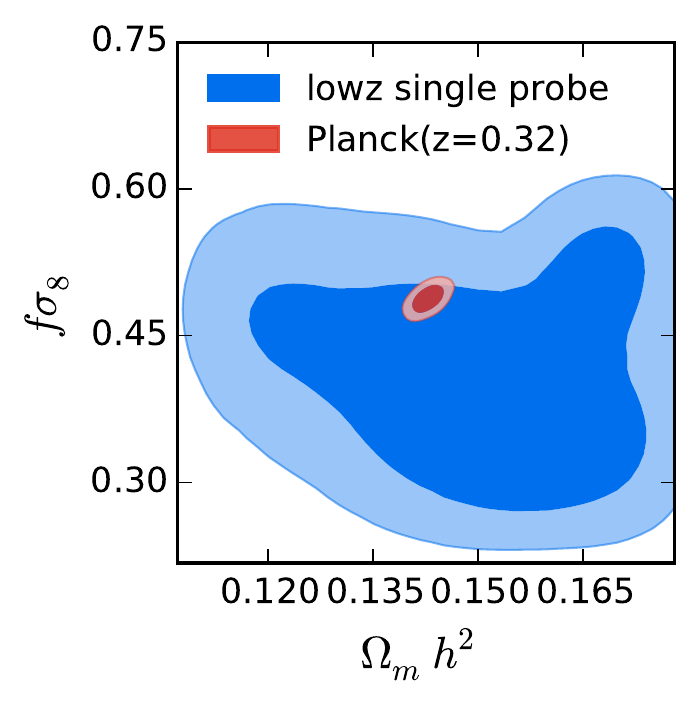}}
\subfigure{\includegraphics[width=1 \columnwidth,clip,angle=0]{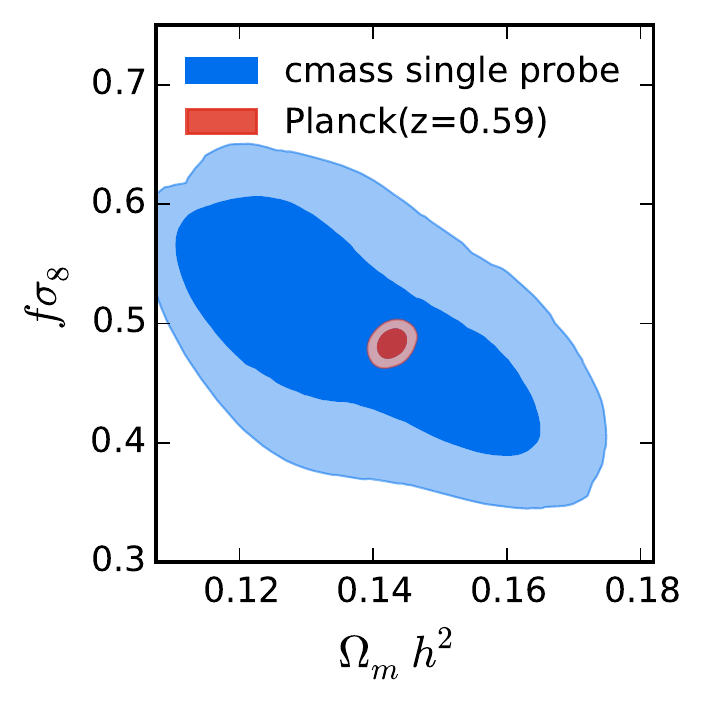}}
\caption{
2D marginalized
  contours for $68\%$ and $95\%$ confidence level for the measurement of $f(z)\sigma_8(z)$ and $\Omega_mh^2$ from the LOWZ sample comparing with the Planck prediction at the same redshift (left panel for $z=0.32$ and right panel for $z=0.59$ ; assuming $\Lambda$CDM).
}
\label{fig:fs8_omh2_planck}
\end{figure*}

There seems to be a slight deviation between our $f(z)\sigma_8(z)$ measurements and Planck $\Lambda$CDM prediction, e.g. in our measurement at $z=0.32$ (Fig. \ref{fig:compare_fs8_lcdm}). In fact, the measurements are consistent with Planck result within 1$\sigma$ if one looks at the 2-dimensional contours of $f(z)\sigma_8(z)$ and $\Omega_mh^2$ shown in Fig. \ref{fig:fs8_omh2_planck}. One can see that there is some correlation between $f(z)\sigma_8(z)$ and $\Omega_mh^2$.

\section{Summary} 
\label{sec:conclusion}
We present measurements of the anisotropic galaxy clustering from the CMASS and LOWZ samples of the final date release (DR12) of the SDSS-III Baryon Oscillation Spectroscopic Survey (BOSS) and obtain constraints on the Hubble expansion rate $H(z)$, the angular-diameter distance $D_A(z)$, 
the normalized growth rate $f(z)\sigma_8(z)$, and the physical matter density $\Omega_mh^2$. 
We analyse the broad-range shape of quasi-linear scales of the monopole and quadrupole correlation functions to obtain cosmological constraints at different redshift bins. In addition to the two redshift bins, i.e. LOWZ ($z_\textrm{LOWZ}=0.32$) and CMASS ($z_\textrm{CMASS}=0.59$), we split each galaxy sample into 2 bins (for a total of 4 redshift bins) and obtain the measurements at $z=\{0.24$, $0.37$, $0.49$, $0.64\}$ to increase the sensitivity of redshift evolution. However, we do not find improvement in terms of constraining different dark energy model parameters. It might indicate that the dark energy component is stable in the redshift range considered. 

We adopt wide and flat priors on all model parameters in order to ensure the results are those of a `single-probe' galaxy clustering analysis. We also marginalize over three nuisance terms that account for potential observational systematics affecting the measured monopole. The Monte Carlo Markov Chain analysis with such wide priors and additional polynomial functions is computationally expensive for advanced theoretical models.
We have developed and validated a new methodology to speed this up by scanning the parameter space using a fast model first and then applying importance sampling using a slower but more accurate model.

Our measurements for DR12 galaxy sample, using the range $40h^{-1}$Mpc $<s<180h^{-1}$Mpc, are
$\{D_A(z)r_{s,fid}/r_s$, $H(z)r_s/r_{s,fid}$, $f(z)\sigma_8(z)$, $\Omega_m h^2\}$ = $\{956\pm28$ $\rm Mpc$, $75.0\pm4.0$ $\Hunit$, $0.397 \pm 0.073$, $0.143\pm0.017\}$ at $z=0.32$ and $\{1421\pm23$ $\rm Mpc$, $96.7\pm2.7$ $\Hunit$, $0.497 \pm 0.058$, $0.137\pm0.015\}$ at $z=0.59$ 
where $r_s$ is the comoving sound horizon at the drag epoch and $r_{s,fid}=147.66$ Mpc is the sound scale of the fiducial cosmology used in this 
study. 
Combining our measurements with Planck data, we obtain 
$\Omega_m=0.306\pm0.009$, $H_0=67.9\pm0.7\Hunit$, and $\sigma_8=0.815\pm0.009$ assuming $\Lambda$CDM; 
$\Omega_k=0.000\pm0.003$ assuming oCDM; $w=-1.01\pm0.06$ assuming $w$CDM; and $w_0=-0.95\pm0.22$ and $w_a=-0.22\pm0.63$ assuming $w_0w_a$CDM. 
The results show no tension with the flat $\Lambda$CDM cosmological paradigm.

\section{Acknowledgements}
C.C. and F.P. acknowledge support from the Spanish MICINN’s Consolider-Ingenio 2010 Programme under grant MultiDark CSD2009-00064 and AYA2010-21231-C02-01 grant.
C.C. was also supported by the Comunidad de Madrid under grant HEPHACOS S2009/ESP-1473. C.C. was supported as a MultiDark fellow.
MPI acknowledges support from MINECO under the grant AYA2012-39702-C02-01.
G.R. is supported by the National Research Foundation of Korea (NRF) through NRF-SGER 2014055950 funded by the Korean Ministry of Education, Science and Technology (MoEST), and by the faculty research fund of Sejong University in 2016.

We acknowledge the use of 
the CURIE supercomputer at Tr\`es Grand Centre de calcul du CEA in France  through the French participation into the PRACE research infrastructure,
the SuperMUC supercomputer at Leibniz Supercomputing Centre of the Bavarian Academy of Science in Germany,
the TEIDE-HPC (High Performance Computing) supercomputer in Spain,
and the Hydra cluster at Instituto de F\'{\i}sica Te\'orica, (UAM/CSIC) in Spain.

Funding for SDSS-III has been provided by the Alfred P. Sloan Foundation, the Participating Institutions, the National Science Foundation, 
and the U.S. Department of Energy Office of Science. The SDSS-III web site is http://www.sdss3.org/.

SDSS-III is managed by the Astrophysical Research Consortium for the Participating Institutions of the SDSS-III Collaboration including 
the University of Arizona, the Brazilian Participation Group, Brookhaven National Laboratory, Carnegie Mellon University, University of Florida, 
the French Participation Group, the German Participation Group, Harvard University, the Instituto de Astrofisica de Canarias, 
the Michigan State/Notre Dame/JINA Participation Group, Johns Hopkins University, Lawrence Berkeley National Laboratory, Max Planck Institute for Astrophysics, 
Max Planck Institute for Extraterrestrial Physics, New Mexico State University, New York University, Ohio State University, 
Pennsylvania State University, University of Portsmouth, Princeton University, the Spanish Participation Group, University of Tokyo, University of Utah, 
Vanderbilt University, University of Virginia, University of Washington, and Yale University. 

\bibliography{singleprobe} 

\appendix
\section{performance of calibrated dewiggle model}\label{sec:table_of_dw}
We present the results using the calibrated dewiggle model in Table \ref{sec:table_of_dw} which also recovers the input parameters with reasonable precision ($0.6\sigma$). It shows that our methodology does not bias significantly our results.

\begin{table*}
 \begin{center}
  \begin{tabular}{c c c c c c} 
     \hline
	&	$\Omega_mh^2$		&		$f\sigma_8(z)$		&		$\frac{H(z)r_s}{r_{s,fid}}$		&		$\frac{D_A(z)r_{s,fid}}{r_s}$		&		$\frac{D_V(z)r_{s,fid}}{r_s}$			\\ \hline
$0.15<z<0.43$	&$	0.150	\pm	0.015	$&$	0.464	\pm	0.086	$&$	79.9	\pm	5.2	$&$	991	\pm	33	$&$	1272	\pm	30	$\\
input values	&$	0.14105			$&$	0.481			$&$	80.16			$&$	990.2			$&$	1269.19			$\\
deviation $\&$ uncertainty ($\%$)	&$	6.3	$ \& $	10.6	$&$	-3.6	$ \& $	18.0	$&$	-0.4	$ \& $	6.5	$&$	0.0	$ \& $	3.3	$&$	0.2	$ \& $	2.3	$\\ \hline
$0.43<z<0.75$	&$	0.150	\pm	0.014	$&$	0.490	\pm	0.055	$&$	93.6	\pm	3.5	$&$	1416	\pm	27	$&$	2124	\pm	30	$\\
input values	&$	0.14105			$&$	0.4786			$&$	94.09			$&$	1409.26			$&$	2113.37			$\\
deviation $\&$ uncertainty ($\%$)	&$	6.2	$ \& $	9.8	$&$	2.3	$ \& $	11.5	$&$	-0.5	$ \& $	3.7	$&$	0.5	$ \& $	1.9	$&$	0.5	$ \& $	1.4	$\\ \hline
$0.15<z<0.30$	&$	0.143	\pm	0.016	$&$	0.469	\pm	0.111	$&$	77.6	\pm	8.8	$&$	802	\pm	58	$&$	973	\pm	56	$\\
input values	&$	0.14105			$&$	0.4751			$&$	76.63			$&$	807.25			$&$	979.874			$\\
deviation $\&$ uncertainty ($\%$)	&$	1.7	$ \& $	11.6	$&$	-1.3	$ \& $	23.3	$&$	1.2	$ \& $	11.4	$&$	-0.7	$ \& $	7.2	$&$	-0.7	$ \& $	5.7	$\\ \hline
$0.30<z<0.43$	&$	0.147	\pm	0.016	$&$	0.489	\pm	0.099	$&$	82.4	\pm	7.1	$&$	1090	\pm	45	$&$	1444	\pm	46	$\\
input values	&$	0.14105			$&$	0.4829			$&$	82.52			$&$	1088.59			$&$	1440.62			$\\
deviation $\&$ uncertainty ($\%$)	&$	3.9	$ \& $	11.2	$&$	1.3	$ \& $	20.6	$&$	-0.1	$ \& $	8.6	$&$	0.1	$ \& $	4.2	$&$	0.2	$ \& $	3.2	$\\ \hline
$0.43<z<0.55$	&$	0.147	\pm	0.016	$&$	0.494	\pm	0.077	$&$	88.0	\pm	5.6	$&$	1287	\pm	37	$&$	1832	\pm	43	$\\
input values	&$	0.14105			$&$	0.4827			$&$	88.59			$&$	1283.41			$&$	1823.53			$\\
deviation $\&$ uncertainty ($\%$)	&$	4.1	$ \& $	11.0	$&$	2.3	$ \& $	15.9	$&$	-0.6	$ \& $	6.4	$&$	0.3	$ \& $	2.9	$&$	0.5	$ \& $	2.3	$\\ \hline
$0.55<z<0.75$	&$	0.145	\pm	0.015	$&$	0.495	\pm	0.071	$&$	97.0	\pm	5.1	$&$	1468	\pm	37	$&$	2255	\pm	44	$\\
input values	&$	0.14105			$&$	0.4754			$&$	96.97			$&$	1461.99			$&$	2248.92			$\\
deviation $\&$ uncertainty ($\%$)	&$	3.0	$ \& $	10.5	$&$	4.1	$ \& $	14.9	$&$	0.0	$ \& $	5.3	$&$	0.4	$ \& $	2.5	$&$	0.3	$ \& $	1.9	$\\ \hline
     \hline
 \end{tabular}
 \end{center}
\caption{Measurements from the mean of 2000 correlation functions using dewiggle model, where the unit of $H(z)$ is $\Hunit$ and the units of $D_A(z)$ and $D_V(z)$ are Mpc.}
\label{table:2000mock_dw}
  \end{table*}

\section{measured covariance matrix}\label{sec:norm_covar}
We show the normalized covariance matrices (also called "correlation matrices") of our measurements in Table \ref{table:cov_z15z43} to \ref{table:cov_z55z75}. A normalized covariance matrix is defined by
\begin{equation}
N_{ij}=\frac{C_{ij}}{\sqrt{C_{ii}C_{jj}}},
\end{equation}
where $C_{ij}$ is the covariance matrix.

 \begin{table*}
 \begin{center}
  \begin{tabular}{c|ccccccc} 
     \hline \\
	&	$D_A(z)\frac{r_{s,fid}}{r_s}$	&	$H(z)\frac{r_s}{r_{s,fid}}$	&	$f\sigma_8(z)$	&	$\Omega_mh^2$	&	$D_V(z)\frac{r_{s,fid}}{r_s}$	&	$\beta(z)$	&	$b\sigma_8(z)$	\\ \hline
$D_A(z)\frac{r_{s,fid}}{r_s}$	&	1.0000	&	0.3899	&	0.5581	&	0.1814	&	0.6045	&	0.4718	&	0.0160	\\
$H(z)\frac{r_s}{r_{s,fid}}$	&	0.3899	&	1.0000	&	0.6710	&	0.3128	&	-0.4973	&	0.6559	&	-0.2362	\\
$f\sigma_8(z)$	&	0.5581	&	0.6710	&	1.0000	&	0.0552	&	-0.0562	&	0.9476	&	-0.2714	\\
$\Omega_mh^2$	&	0.1814	&	0.3128	&	0.0552	&	1.0000	&	-0.1057	&	0.0364	&	0.0756	\\
$D_V(z)\frac{r_{s,fid}}{r_s}$	&	0.6045	&	-0.4973	&	-0.0562	&	-0.1057	&	1.0000	&	-0.1237	&	0.2183	\\
$\beta(z)$	&	0.4718	&	0.6559	&	0.9476	&	0.0364	&	-0.1237	&	1.0000	&	-0.5544	\\
$b\sigma_8(z)$	&	0.0160	&	-0.2362	&	-0.2714	&	0.0756	&	0.2183	&	-0.5544	&	1.0000	\\
     \hline
 \end{tabular}
 \end{center}
\caption{Normalized covariance matrix of the measurements from DR12 galaxy sample of $0.15 < z < 0.43$. the units of $D_A(z)$ and $D_V(z)$ are Mpc.}
\label{table:cov_z15z43}
  \end{table*}

 \begin{table*}
 \begin{center}
  \begin{tabular}{c|c c c c c c c} 
     \hline \\
	&	$D_A(z)\frac{r_{s,fid}}{r_s}$	&	$H(z)\frac{r_s}{r_{s,fid}}$	&	$f\sigma_8(z)$	&	$\Omega_mh^2$	&	$D_V(z)\frac{r_{s,fid}}{r_s}$	&	$\beta(z)$	&	$b\sigma_8(z)$	\\ \hline
$D_A(z)\frac{r_{s,fid}}{r_s}$	&	1.0000	&	0.4402	&	0.5222	&	-0.0847	&	0.6206	&	0.4513	&	-0.1722	\\
$H(z)\frac{r_s}{r_{s,fid}}$	&	0.4402	&	1.0000	&	0.5410	&	-0.2255	&	-0.4306	&	0.4293	&	-0.0799	\\
$f\sigma_8(z)$	&	0.5222	&	0.5410	&	1.0000	&	-0.5879	&	0.0509	&	0.8951	&	-0.3739	\\
$\Omega_mh^2$	&	-0.0847	&	-0.2255	&	-0.5879	&	1.0000	&	0.1152	&	-0.5034	&	0.1335	\\
$D_V(z)\frac{r_{s,fid}}{r_s}$	&	0.6206	&	-0.4306	&	0.0509	&	0.1152	&	1.0000	&	0.0769	&	-0.1022	\\
$\beta(z)$	&	0.4513	&	0.4293	&	0.8951	&	-0.5034	&	0.0769	&	1.0000	&	-0.7402	\\
$b\sigma_8(z)$	&	-0.1722	&	-0.0799	&	-0.3739	&	0.1335	&	-0.1022	&	-0.7402	&	1.0000	\\
     \hline
 \end{tabular}
 \end{center}
\caption{Normalized covariance matrix of the measurements from DR12 galaxy sample of $0.43 < z < 0.75$. the units of $D_A(z)$ and $D_V(z)$ are Mpc.}
\label{table:cov_z43z75}
  \end{table*}

 \begin{table*}
 \begin{center}
  \begin{tabular}{c|c c c c c c c} 
     \hline \\
	&	$D_A(z)\frac{r_{s,fid}}{r_s}$	&	$H(z)\frac{r_s}{r_{s,fid}}$	&	$f\sigma_8(z)$	&	$\Omega_mh^2$	&	$D_V(z)\frac{r_{s,fid}}{r_s}$	&	$\beta(z)$	&	$b\sigma_8(z)$	\\ \hline
$D_A(z)\frac{r_{s,fid}}{r_s}$	&	1.0000	&	0.1492	&	0.5334	&	0.0738	&	0.8167	&	0.3523	&	0.2295	\\
$H(z)\frac{r_s}{r_{s,fid}}$	&	0.1492	&	1.0000	&	0.4036	&	0.1066	&	-0.4465	&	0.4294	&	-0.2197	\\
$f\sigma_8(z)$	&	0.5334	&	0.4036	&	1.0000	&	-0.1680	&	0.2417	&	0.9096	&	-0.1677	\\
$\Omega_mh^2$	&	0.0738	&	0.1066	&	-0.1680	&	1.0000	&	0.0131	&	-0.1872	&	0.0796	\\
$D_V(z)\frac{r_{s,fid}}{r_s}$	&	0.8167	&	-0.4465	&	0.2417	&	0.0131	&	1.0000	&	0.0625	&	0.3418	\\
$\beta(z)$	&	0.3523	&	0.4294	&	0.9096	&	-0.1872	&	0.0625	&	1.0000	&	-0.5489	\\
$b\sigma_8(z)$	&	0.2295	&	-0.2197	&	-0.1677	&	0.0796	&	0.3418	&	-0.5489	&	1.0000	\\
     \hline
 \end{tabular}
 \end{center}
\caption{Normalized covariance matrix of the measurements from DR12 galaxy sample of $0.15 < z < 0.30$. the units of $D_A(z)$ and $D_V(z)$ are Mpc.}
\label{table:cov_z15z30}
  \end{table*}

 \begin{table*}
 \begin{center}
  \begin{tabular}{c|c c c c c c c} 
     \hline \\
	&	$D_A(z)\frac{r_{s,fid}}{r_s}$	&	$H(z)\frac{r_s}{r_{s,fid}}$	&	$f\sigma_8(z)$	&	$\Omega_mh^2$	&	$D_V(z)\frac{r_{s,fid}}{r_s}$	&	$\beta(z)$	&	$b\sigma_8(z)$	\\ \hline
$D_A(z)\frac{r_{s,fid}}{r_s}$	&	1.0000	&	0.1042	&	0.5015	&	0.1169	&	0.8364	&	0.1788	&	0.5662	\\
$H(z)\frac{r_s}{r_{s,fid}}$	&	0.1042	&	1.0000	&	0.4615	&	-0.1769	&	-0.4533	&	0.5100	&	-0.1488	\\
$f\sigma_8(z)$	&	0.5015	&	0.4615	&	1.0000	&	-0.2777	&	0.1991	&	0.8736	&	0.0567	\\
$\Omega_mh^2$	&	0.1169	&	-0.1769	&	-0.2777	&	1.0000	&	0.2003	&	-0.2214	&	-0.0643	\\
$D_V(z)\frac{r_{s,fid}}{r_s}$	&	0.8364	&	-0.4533	&	0.1991	&	0.2003	&	1.0000	&	-0.1108	&	0.5839	\\
$\beta(z)$	&	0.1788	&	0.5100	&	0.8736	&	-0.2214	&	-0.1108	&	1.0000	&	-0.4223	\\
$b\sigma_8(z)$	&	0.5662	&	-0.1488	&	0.0567	&	-0.0643	&	0.5839	&	-0.4223	&	1.0000	\\
     \hline
 \end{tabular}
 \end{center}
\caption{Normalized covariance matrix of the measurements from DR12 galaxy sample of $0.30 < z < 0.43$. the units of $D_A(z)$ and $D_V(z)$ are Mpc.}
\label{table:cov_z30z43}
  \end{table*}

 \begin{table*}
 \begin{center}
  \begin{tabular}{c|c c c c c c c} 
     \hline \\
	&	$D_A(z)\frac{r_{s,fid}}{r_s}$	&	$H(z)\frac{r_s}{r_{s,fid}}$	&	$f\sigma_8(z)$	&	$\Omega_mh^2$	&	$D_V(z)\frac{r_{s,fid}}{r_s}$	&	$\beta(z)$	&	$b\sigma_8(z)$	\\ \hline
$D_A(z)\frac{r_{s,fid}}{r_s}$	&	1.0000	&	0.3189	&	0.4258	&	0.0776	&	0.5088	&	0.2947	&	0.0573	\\
$H(z)\frac{r_s}{r_{s,fid}}$	&	0.3189	&	1.0000	&	0.5740	&	0.0618	&	-0.6525	&	0.5351	&	-0.2326	\\
$f\sigma_8(z)$	&	0.4258	&	0.5740	&	1.0000	&	-0.2981	&	-0.1848	&	0.9001	&	-0.3283	\\
$\Omega_mh^2$	&	0.0776	&	0.0618	&	-0.2981	&	1.0000	&	0.0109	&	-0.1441	&	-0.1903	\\
$D_V(z)\frac{r_{s,fid}}{r_s}$	&	0.5088	&	-0.6525	&	-0.1848	&	0.0109	&	1.0000	&	-0.2533	&	0.2576	\\
$\beta(z)$	&	0.2947	&	0.5351	&	0.9001	&	-0.1441	&	-0.2533	&	1.0000	&	-0.6958	\\
$b\sigma_8(z)$	&	0.0573	&	-0.2326	&	-0.3283	&	-0.1903	&	0.2576	&	-0.6958	&	1.0000	\\
     \hline
 \end{tabular}
 \end{center}
\caption{Normalized covariance matrix of the measurements from DR12 galaxy sample of $0.43 < z < 0.55$. the units of $D_A(z)$ and $D_V(z)$ are Mpc.}
\label{table:cov_z43z55}
  \end{table*}

 \begin{table*}
 \begin{center}
  \begin{tabular}{c|c c c c c c c} 
     \hline \\
	&	$D_A(z)\frac{r_{s,fid}}{r_s}$	&	$H(z)\frac{r_s}{r_{s,fid}}$	&	$f\sigma_8(z)$	&	$\Omega_mh^2$	&	$D_V(z)\frac{r_{s,fid}}{r_s}$	&	$\beta(z)$	&	$b\sigma_8(z)$	\\ \hline
$D_A(z)\frac{r_{s,fid}}{r_s}$	&	1.0000	&	0.4299	&	0.4490	&	0.0544	&	0.7736	&	0.3330	&	0.0055	\\
$H(z)\frac{r_s}{r_{s,fid}}$	&	0.4299	&	1.0000	&	0.4408	&	-0.0347	&	-0.2390	&	0.3523	&	-0.0624	\\
$f\sigma_8(z)$	&	0.4490	&	0.4408	&	1.0000	&	-0.4533	&	0.1753	&	0.8950	&	-0.3339	\\
$\Omega_mh^2$	&	0.0544	&	-0.0347	&	-0.4533	&	1.0000	&	0.0815	&	-0.3508	&	0.0318	\\
$D_V(z)\frac{r_{s,fid}}{r_s}$	&	0.7736	&	-0.2390	&	0.1753	&	0.0815	&	1.0000	&	0.1114	&	0.0513	\\
$\beta(z)$	&	0.3330	&	0.3523	&	0.8950	&	-0.3508	&	0.1114	&	1.0000	&	-0.7106	\\
$b\sigma_8(z)$	&	0.0055	&	-0.0624	&	-0.3339	&	0.0318	&	0.0513	&	-0.7106	&	1.0000	\\
     \hline
 \end{tabular}
 \end{center}
\caption{Normalized covariance matrix of the measurements from DR12 galaxy sample of $0.55 < z < 0.75$. the units of $D_A(z)$ and $D_V(z)$ are Mpc.}
\label{table:cov_z55z75}
  \end{table*}

\section{compilations of measurements from other works and this study}\label{sec:compile_previous}
We compile the measurements of $f(z)\sigma_8(z)$, $D_A(z)/r_s$, $H(z)*r_s$, and $D_V(z)/r_s$ from various galaxy surveys in Table \ref{table:fs8}, \ref{table:DVrs}, and \ref{table:H_DA}. 
We have included the measurements from 
VIMOS-VLT Deep Survey (VVDS; \citealt{Guzzo:2008ac}),
2dFGRS \citep{Percival:2004fs},
Six-degree-Field Galaxy Survey (6dFGS; \citealt{Beutler:2011hx,Beutler:2012px}),
WiggleZ \citep{Blake:2011en,Blake:2011rj,Blake:2012pj,Contreras:2013bol},
SDSS-II/DR7 \citep{Percival:2009xn,Chuang:2010dv,Samushia:2011cs,Chuang:2011fy,Chuang:2012ad,Chuang:2012qt,Ross:2014qpa,Padmanabhan:2012hf,Xu:2012fw, 
Seo:2012xy,Hemantha:2013sea}
SDSS-III/BOSS \citep{Anderson:2012sa,Reid:2012sw, Anderson:2013oza, Chuang:2013hya, Kazin:2013rxa,Wang:2014qoa, Anderson:2013zyy,Beutler:2013yhm,Samushia:2013yga,Sanchez:2013tga,Tojeiro:2014eea,Reid:2014iaa,Alam:2015qta,Gil-Marin:2015nqa,Gil-Marin:2015sqa,Cuesta:2015mqa},
\cite{Acacia} (BOSS collaboration paper for final data release) and its companion papers including this paper and 
\cite{Ross16, 
Vargas-Magana16, 
Beutler16b, 
Satpathy16, 
Beutler16c, 
Sanchez16a, 
Grieb16, 
Pellejero-Ibanez16}.
To be able to include more measurements, we quote $D_V(z)/r_s$  instead of $D_V(z)r_{s,fid}/r_s$ since  $r_{s,fid}$ was not provided in some references.

\begin{table*}
\begin{center}
\begin{tabular}{cccc}
\hline
Redshift	&	$f(z)\sigma_8(z)$			&	Data	&Reference	\\	\hline
0.64	&$	0.454	\pm	0.064	$&	DR12	&This study	\\	
0.59	&$	0.497	\pm	0.058	$&	DR12	& 	\\	
0.49	&$	0.456	\pm	0.068	$&	DR12	& 	\\	
0.37	&$	0.378	\pm	0.076	$&	DR12	& 	\\	
0.32	&$	0.397	\pm	0.073	$&	DR12	& 	\\	
0.24	&$	0.493	\pm	0.105	$&	DR12	& 	\\	\hline
0.59	&$	0.51	\pm	0.047	$&	DR12	&\cite{Pellejero-Ibanez16}\\	
0.32	&$	0.431	\pm	0.063	$&	DR12	& 	\\	\hline
0.61	&$	0.436	\pm	0.034	$&	DR12	&\cite{Acacia}	\\	
0.51	&$	0.458	\pm	0.035	$&	DR12	&(BOSS consensus results) 	\\	
0.38	&$	0.497	\pm	0.039	$&	DR12	& 	\\	\hline
0.61	&$	0.456	\pm	0.052	$&	DR12	&\cite{Satpathy16}\\	
0.51	&$	0.452	\pm	0.058	$&	DR12	& 	\\	
0.38	&$	0.43	\pm	0.054	$&	DR12	& 	\\	\hline
0.61	&$	0.409	\pm	0.044	$&	DR12	&\cite{Beutler16b}	\\	
0.51	&$	0.454	\pm	0.051	$&	DR12	& 	\\	
0.38	&$	0.479	\pm	0.054	$&	DR12	& 	\\	\hline
0.61	&$	0.409	\pm	0.041	$&	DR12	&\cite{Grieb16}	\\	
0.51	&$	0.448	\pm	0.038	$&	DR12	& 	\\	
0.38	&$	0.498	\pm	0.045	$&	DR12	& 	\\	\hline
0.61	&$	0.44	\pm	0.039	$&	DR12	&\cite{Sanchez16a}\\	
0.51	&$	0.47	\pm	0.042	$&	DR12	& 	\\	
0.38	&$	0.468	\pm	0.053	$&	DR12	& 	\\	\hline
0.59	&$	0.488	\pm	0.06	$&	DR12	&\cite{Chuang:2013wga}	\\	\hline
0.57	&$	0.444	\pm	0.038	$&	DR12	&\cite{Gil-Marin:2015sqa}	\\	
0.32	&$	0.394	\pm	0.062	$&	DR12	& 	\\	\hline
0.57	&$	0.417	\pm	0.045	$&	DR11	&\cite{Sanchez:2013tga}	\\	
0.32	&$	0.48	\pm	0.10	$&	DR11	& 	\\	\hline
0.57	&$	0.441	\pm	0.044	$&	DR11	&\cite{Samushia:2013yga}	\\	
0.57	&$	0.419	\pm	0.044	$&	DR11	&\cite{Beutler:2013yhm}	\\	
0.57	&$	0.462	\pm	0.041	$&	DR11	&\cite{Alam:2015qta}	\\	
0.57	&$	0.45	\pm	0.011	$&	DR10	&\cite{Reid:2014iaa}	\\	
0.57	&$	0.428	\pm	0.069	$&	DR9	&\cite{Chuang:2013hya}	\\	
0.57	&$	0.415	\pm	0.034	$&	DR9	&\cite{Reid:2012sw}	\\	
0.57	&$	0.474	\pm	0.075	$&	DR9	&\cite{Wang:2014qoa}	\\	
0.3	&$	0.49	\pm	0.08	$&	DR7	&\cite{Oka:2013cba}	\\	\hline
0.37	&$	0.46	\pm	0.04	$&	DR7	&\cite{Samushia:2011cs}	\\	
0.25	&$	0.35	\pm	0.06	$&	DR7	& 	\\	\hline
0.35	&$	0.429	\pm	0.089	$&	DR7	&\cite{Chuang:2012qt}	\\	
0.067	&$	0.423	\pm	0.055	$&	6dFGS	&\cite{Beutler:2012px}	\\	\hline
0.44	&$	0.413	\pm	0.08	$&	WiggleZ	&\cite{Blake:2012pj}	\\	
0.6	&$	0.39	\pm	0.063	$&	WiggleZ	& 	\\	
0.73	&$	0.437	\pm	0.072	$&	WiggleZ	& 	\\	
0.22	&$	0.42	\pm	0.07	$&	WiggleZ	&\cite{Blake:2011rj}	\\	
0.41	&$	0.45	\pm	0.04	$&	WiggleZ	& 	\\	
0.6	&$	0.43	\pm	0.04	$&	WiggleZ	& 	\\	
0.78	&$	0.38	\pm	0.04	$&	WiggleZ	& 	\\	\hline
0.17	&$	0.51	\pm	0.06	$&	2dFGRS	&\cite{Percival:2004fs}	\\	
0.77	&$	0.49	\pm	0.18	$&	VVDS	&\cite{Guzzo:2008ac}	\\	
\hline
\end{tabular}
\end{center}
\caption{ 
Measurements of $f(z)\sigma_8(z)$ from different galaxy surveys, including SDSS-II (DR7), SDSS-III (DR9,DR11,DR12), 6dFGS, WiggleZ, 2dFGRS, and VVDS.
} \label{table:fs8}
\end{table*}

\begin{table*}
\begin{center}
\begin{tabular}{ccccc}
\hline
Redshift	&	$D_V(z)/r_s$			&	$r_{s,fid}$	&	Data	&Reference	\\	\hline
0.64	&$	15.03	\pm	0.26	$&	147.66	&	DR12	&This study	\\	
0.59	&$	14.26	\pm	0.16	$&	147.66	&	DR12	& 	\\	
0.49	&$	12.44	\pm	0.24	$&	147.66	&	DR12	& 	\\	
0.37	&$	9.49	\pm	0.47	$&	147.66	&	DR12	& 	\\	
0.32	&$	8.59	\pm	0.18	$&	147.66	&	DR12	& 	\\	
0.24	&$	6.68	\pm	0.27	$&	147.66	&	DR12	& 	\\	\hline
0.61	&$	14.48	\pm	0.15	$&	147.78	&	DR12	&\cite{Acacia}	\\	
0.51	&$	12.70	\pm	0.13	$&	147.78	&	DR12	&(BOSS consensus results)	\\	
0.38	&$	9.99	\pm	0.11	$&	147.78	&	DR12	&	\\	\hline
0.59	&$	14.27	\pm	0.18	$&	147.66	&	DR12	&\cite{Chuang:2013wga}	\\	\hline
0.57	&$	13.79	\pm	0.14	$&	147.1	&	DR12	&\cite{Cuesta:2015mqa}	\\	
0.32	&$	8.59	\pm	0.15	$&	147.1	&	DR12	& 	\\	\hline
0.57	&$	13.70	\pm	0.12	$&		&	DR12	&\cite{Gil-Marin:2015nqa}	\\	
0.32	&$	8.62	\pm	0.15	$&		&	DR12	& 	\\	\hline
0.57	&$	13.85	\pm	0.17	$&		&	DR11	&\cite{Samushia:2013yga}	\\	
0.57	&$	13.89	\pm	0.18	$&	147.36	&	DR11	&\cite{Beutler:2013yhm}	\\	\hline
0.57	&$	13.77	\pm	0.13	$&	149.28	&	DR11	&\cite{Anderson:2013zyy}	\\	
0.32	&$	8.47	\pm	0.17	$&	149.28	&	DR11	& 	\\	\hline
0.32	&$	8.47	\pm	0.17	$&	149.28	&	DR11	&\cite{Tojeiro:2014eea}	\\	
0.57	&$	14.04	\pm	0.23	$&	149.16	&	DR9	&\cite{Anderson:2012sa}	\\	
0.57	&$	13.91	\pm	0.30	$&		&	DR9	&\cite{Chuang:2013hya}	\\	
0.35	&$	9.12	\pm	0.17	$&		&	DR7	&\cite{Padmanabhan:2012hf}	\\	
0.35	&$	8.85	\pm	0.26	$&		&	DR7	&\cite{Chuang:2011fy}	\\	
0.35	&$	8.99	\pm	0.24	$&		&	DR7	&\cite{Chuang:2010dv}	\\	\hline
0.35	&$	9.37	\pm	0.31	$&		&	DR7+2dFGRS	&\cite{Percival:2009xn}	\\	
0.2	&$	5.39	\pm	0.17	$&		&	DR7+2dFGRS	& 	\\	\hline
0.15	&$	4.47	\pm	0.17	$&	148.69	&	DR7	&\cite{Ross:2014qpa}	\\	
0.106	&$	3.06	\pm	0.13	$&		&	6dFGS	&\cite{Beutler:2011hx}	\\	\hline
0.44	&$	11.50	\pm	0.56	$&	149.28	&	WiggleZ	&\cite{Kazin:2014qga}	\\	
0.6	&$	14.88	\pm	0.68	$&	149.28	&	WiggleZ	& 	\\	
0.73	&$	16.85	\pm	0.58	$&	149.28	&	WiggleZ	& 	\\	
0.44	&$	11.20	\pm	0.87	$&		&	WiggleZ	&\cite{Blake:2011en}	\\	
0.6	&$	14.14	\pm	0.67	$&		&	WiggleZ	& 	\\	
0.73	&$	17.35	\pm	0.93	$&		&	WiggleZ	& 	\\	
\hline
\end{tabular}
\end{center}
\caption{ 
Measurements of  $D_V(z)/r_s$ from different galaxy surveys, including SDSS-II (DR7), SDSS-III (DR9,DR11,DR12), 2dFGRS, 6dFGS, and WiggleZ. To be able to include more measurements, we use $D_V(z)/r_s$ instead of $D_V(z)r_{s,fid}/r_s$ since  $r_{s,fid}$ (Mpc) was not provided in some literatures. In addition, we have included an approximation $r_{s,\rm EH}/r_{s,\rm CAMB}=1.027$ to correct the different ways of estimating the sound horizon in different analyses.
} \label{table:DVrs}
\end{table*}

\begin{table*}
\begin{center}
\begin{tabular}{ccccccc}
\hline
Redshift	&	$H(z)r_s$			&	$D_A(z)/r_s$			&	$r_{s,fid}$	&	Data	&Reference	\\	\hline
0.64	&$	14530	\pm	546	$&$	9.78	\pm	0.28	$&	147.66	&	DR12	&This study	\\	
0.59	&$	14279	\pm	399	$&$	9.62	\pm	0.16	$&	147.66	&	DR12	& 	\\	
0.49	&$	12920	\pm	709	$&$	8.72	\pm	0.21	$&	147.66	&	DR12	& 	\\	
0.37	&$	11045	\pm	930	$&$	6.72	\pm	0.44	$&	147.66	&	DR12	& 	\\	
0.32	&$	11075	\pm	591	$&$	6.47	\pm	0.19	$&	147.66	&	DR12	& 	\\	
0.24	&$	11636	\pm	827	$&$	5.59	\pm	0.30	$&	147.66	&	DR12	& 	\\	\hline
0.59	&$	14456	\pm	458	$&$	9.63	\pm	0.17	$&	147.66	&	DR12	&\cite{Pellejero-Ibanez16}	\\	
0.32	&$	11680	\pm	487	$&$	6.47	\pm	0.18	$&	147.66	&	DR12	& 	\\	\hline
0.61	&$	14379	\pm	266	$&$	9.60	\pm	0.13	$&	147.78	&	DR12	&\cite{Acacia}	\\	
0.51	&$	13374	\pm	251	$&$	8.86	\pm	0.12	$&	147.78	&	DR12	& (BOSS consensus results)	\\	
0.38	&$	12044	\pm	281	$&$	7.44	\pm	0.11	$&	147.78	&	DR12	& 	\\	\hline
0.61	&$	14601	\pm	340	$&$	9.70	\pm	0.17	$&	147.78	&	DR12	&\cite{Beutler16b}	\\	
0.51	&$	13418	\pm	325	$&$	8.86	\pm	0.14	$&	147.78	&	DR12	& 	\\	
0.38	&$	11926	\pm	355	$&$	7.39	\pm	0.12	$&	147.78	&	DR12	& 	\\	\hline
0.61	&$	14675	\pm	369	$&$	9.63	\pm	0.18	$&	147.78	&	DR12	&\cite{Vargas-Magana16}	\\	
0.51	&$	13448	\pm	310	$&$	8.85	\pm	0.13	$&	147.78	&	DR12	& 	\\	
0.38	&$	11882	\pm	355	$&$	7.39	\pm	0.11	$&	147.78	&	DR12	& 	\\	\hline
0.61	&$	14689	\pm	325	$&$	9.65	\pm	0.18	$&	147.78	&	DR12	&\cite{Ross16}	\\	
0.51	&$	13463	\pm	310	$&$	8.83	\pm	0.13	$&	147.78	&	DR12	& 	\\	
0.38	&$	11985	\pm	325	$&$	7.41	\pm	0.11	$&	147.78	&	DR12	& 	\\	\hline
0.61	&$	14704	\pm	649	$&$	9.61	\pm	0.26	$&	147.78	&	DR12	&\cite{Satpathy16}	\\	
0.51	&$	13053	\pm	607	$&$	8.89	\pm	0.20	$&	147.78	&	DR12	& 	\\	
0.38	&$	11716	\pm	480	$&$	7.24	\pm	0.16	$&	147.78	&	DR12	& 	\\	\hline
0.61	&$	14330	\pm	591	$&$	9.54	\pm	0.28	$&	147.78	&	DR12	&\cite{Beutler16c}\\	
0.51	&$	13064	\pm	599	$&$	9.03	\pm	0.26	$&	147.78	&	DR12	& 	\\	
0.38	&$	12193	\pm	474	$&$	7.59	\pm	0.20	$&	147.78	&	DR12	& 	\\	\hline
0.61	&$	14021	\pm	375	$&$	9.59	\pm	0.21	$&	147.78	&	DR12	&\cite{Grieb16}	\\	
0.51	&$	12863	\pm	349	$&$	8.92	\pm	0.16	$&	147.78	&	DR12	& 	\\	
0.38	&$	11995	\pm	337	$&$	7.48	\pm	0.12	$&	147.78	&	DR12	& 	\\	\hline
0.61	&$	14378	\pm	400	$&$	9.61	\pm	0.18	$&	147.78	&	DR12	&\cite{Sanchez16a}	\\	
0.51	&$	13334	\pm	364	$&$	9.01	\pm	0.15	$&	147.78	&	DR12	& 	\\	
0.38	&$	12186	\pm	352	$&$	7.36	\pm	0.13	$&	147.78	&	DR12	& 	\\	\hline
0.59	&$	14367	\pm	487	$&$	9.66	\pm	0.18	$&	147.66	&	DR12	&\cite{Chuang:2013wga}	\\	\hline
0.57	&$	14754	\pm	544	$&$	9.52	\pm	0.14	$&	147.1	&	DR12	&\cite{Cuesta:2015mqa}	\\	
0.32	&$	11650	\pm	824	$&$	6.67	\pm	0.14	$&	147.1	&	DR12	& 	\\	\hline
0.57	&$	13920	\pm	440	$&$	9.42	\pm	0.15	$&		&	DR12	&\cite{Gil-Marin:2015sqa}	\\	
0.32	&$	11410	\pm	560	$&$	6.35	\pm	0.19	$&		&	DR12	& 	\\	\hline
0.57	&$	14560	\pm	370	$&$	9.42	\pm	0.13	$&		&	DR12	&\cite{Gil-Marin:2015nqa}	\\	
0.32	&$	11600	\pm	600	$&$	6.66	\pm	0.16	$&		&	DR12	& 	\\	\hline
0.57	&$	13719	\pm	486	$&$	9.42	\pm	0.15	$&	147.36	&	DR11	&\cite{Beutler:2013yhm}	\\	
0.57	&$	13960	\pm	448	$&$	9.26	\pm	0.17	$&		&	DR11	&\cite{Sanchez:2013tga}	\\	
0.32	&$	12199	\pm	627	$&$	6.46	\pm	0.28	$&		&	DR11	& 	\\	\hline
0.57	&$	14450	\pm	508	$&$	9.52	\pm	0.13	$&	149.28	&	DR11	&\cite{Anderson:2013zyy}	\\	
0.57	&$	13857	\pm	1163	$&$	9.44	\pm	0.30	$&	149.16	&	DR9	&\cite{Anderson:2013oza}	\\	
0.57	&$	13564	\pm	906	$&$	9.29	\pm	0.28	$&		&	DR9	&\cite{Kazin:2013rxa}	\\	
0.57	&$	13262	\pm	906	$&$	9.19	\pm	0.28	$&		&	DR9	&\cite{Chuang:2013hya}	\\	
0.57	&$	12970	\pm	555	$&$	9.25	\pm	0.24	$&		&	DR9	&\cite{Wang:2014qoa}	\\	
0.35	&$	12590	\pm	526	$&$	6.65	\pm	0.26	$&		&	DR7	&\cite{Hemantha:2013sea}	\\	
0.35	&$	12556	\pm	1042	$&$	7.07	\pm	0.26	$&		&	DR7	&\cite{Xu:2012fw}	\\	
0.35	&$	12648	\pm	1227	$&$	6.77	\pm	0.47	$&		&	DR7	&\cite{Chuang:2012qt}	\\	
0.35	&$	12765	\pm	1227	$&$	6.65	\pm	0.45	$&		&	DR7	&\cite{Chuang:2012ad}	\\	
0.35	&$	12678	\pm	526	$&$	6.78	\pm	0.27	$&		&	DR7	&\cite{Chuang:2011fy}	\\	
\hline
\end{tabular}
\end{center}
\caption{ 
Measurements of $H(z)r_s$ (km/s) and $D_A(z)/r_s$ from different galaxy surveys, including SDSS-II (DR7) and SDSS-III (DR9,DR11,DR12). To be able to include more measurements, we use $D_A(z)/r_s$ and $H(z)r_s$ instead of $D_A(z)r_{s,fid}/r_s$ and $H(z)r_s/r_{s,fid}$ since $r_{s,fid}$ (Mpc) was not provided in some literatures. In addition, we have included an approximation $r_{s,\rm EH}/r_{s,\rm CAMB}=1.027$ to correct the different ways of estimating the sound horizon in different analyses.
} \label{table:H_DA}
\end{table*}

\label{lastpage}

\end{document}